\documentclass[a4paper]{eccomas_paper-2024}
\usepackage{graphicx}

\usepackage{hyperref}
\usepackage{array}  
\usepackage{graphicx}
\usepackage{amsmath}
\usepackage{amsfonts}
\usepackage{amssymb}
\usepackage{url}
\usepackage{graphicx}
\usepackage{tikz}
\usetikzlibrary{shapes,arrows}
\usepackage{pgfplots}
\pgfplotsset{every axis/.append style={thick}}
\pgfplotsset{compat=1.17}
\pgfplotsset{every axis legend/.append
  style={cells={anchor=west},anchor=west}}
\pgfplotscreateplotcyclelist{mylines}{%
  {red,mark=*},
  {blue,mark=square},
  {green,mark=+},
  {black,mark=o},
  {violet,mark=otimes},
  {brown,mark=triangle},
  {cyan,mark=diamond},
  {orange,mark=10-pointed star},
  {magenta,mark=|}  }

\title{Isogeometric Analysis for 2D Magnetostatic Computations with Multi-level B\'{e}zier Extraction for Local Refinement}

\author{Andreas Grendas$^{1}$, Michael Wiesheu$^{2}$, Sebastian Schöps$^{2}$ and Benjamin Marussig$^{1}$}

\heading{A. Grendas, M. Wiesheu, S. Schöps, and B. Marussig}

\address{$^{1}$ Graz University of Technology \\
Institute of Applied Mechanics \\ 
Technikerstraße 4/II, 8010 Graz, Austria \\
a.grendas@tugraz.at, marussig@tugraz.at \\
www.tugraz.at/institute/am-bm
\and
$^{2}$ Computational Electromagnetics Group \\
Technical University of Darmstadt \\
Schloßgartenstr. 8, 64289  Darmstadt, Germany\\
michael.wiesheu@tu-darmstadt.de, sebastian.schoeps@tu-darmstadt.de\\ 
www.cem.tu-darmstadt.de/cem/start/index.en.jsp
}


\abstract{Local refinement is vital for efficient numerical simulations. In the context of Isogeometric Analysis (IGA), hierarchical B-splines have gained prominence. The work applies the methodology of truncated hierarchical B-splines (THB-splines) as they keep additional properties. The framework is further enriched with B\'{e}zier extraction, resulting in the multi-level B\'{e}zier extraction method. We apply this discretization method to 2D magnetostatic problems. The implementation is based on an open-source Octave/MATLAB IGA code called GeoPDEs, which allows us to compare our routines with globally refined spline models as well as locally refined ones where the solver does not rely on B\'{e}zier extraction.}

\begin{document}
\thispagestyle{empty}

\section{Introduction}

In IGA \cite{IGA_hughes}, the Basis splines (B-spline) and Non-Uniform Rational B-splines (NURBS)  functions used to represent the CAD geometry are also employed to approximate the field variables within the numerical model. This approach reduces the geometric discretization errors inherent in finite element models that use an approximated computational geometry.

A key topic is the generalization of spline constructions, enabling local refinement by breaking the global tensor product structure of multivariate splines. Several methods have been devised to address this limitation of tensor product structures, such as T-splines \cite{T-splines}, locally refined B-spline (LR-splines) \cite{LR}, and hierarchical B-splines (HB-splines) \cite{HB}. HB-splines operate on a multi-level structure with a sequence of different levels of detail and perform local refinement by activating and deactivating of B-splines on these levels. The HB-splines can be extended with a truncation mechanism, which leads to truncated HB-splines (THB-splines) \cite{THB}. Additional information about the THB-splines will be discussed in the corresponding section.

This work introduces an adaptive IGA framework based on THB-splines, which is extended with the discretization technique of classical B\'{e}zier extraction \cite{BE}. This approach could potentially implement the hierarchical B-spline shape functions into an existing finite element (FE) solver by utilizing B\'{e}zier element with a standard single-level basis (Bernstein polynomials), in other words, multi-level B\'{e}zier extraction \cite{MBE1,MBE2}.

Our implementation relies on GeoPDEs, an open-source Octave/MATLAB IGA code \cite{GeoPDEs}, where we can provide proper comparisons as the tool already offers adaptivity based on THB-splines. For the numerical example, we solve a scalar 2D magnetostatic problem. In addition, a posteriori error estimator is introduced, which requires further investigation for more complex geometries, but it can promise refinement without an exact solution.

\section{Background}

This section summarizes the relevant concepts to offer a clear overview of how the adaptivity method is employed in IGA by utilizing multi-level B\'{e}zier extraction. Starting with B-splines, followed by the multi-level structure based on THB-splines, and finally applying the B\'{e}zier extraction technique.

\subsection{B-splines}

In IGA, B-splines serve as the basis functions for representing geometry and performing the simulation analysis \cite{IGA_hughes,nurbs}. B-spline basis functions are defined by degree $p$, and a knot vector,
\begin{eqnarray}\label{one}
     \Xi &=& \{ \xi_1\leq...\leq \xi_{n+p+1} \},
\end{eqnarray}
where $n$ is the number of the basis functions. The Cox-de Boor recursion formula defines the $n$ functions, by starting from degree $p=0$, up to the prescribed degree, 

\begin{eqnarray}\label{Cox}
    N_{i,0} &=& \begin{cases}
                    1 ~ \text {for} ~ \xi_i ~ \leq \xi < \xi_{i+1}  \\
                    0 ~ \text {else}
                 \end{cases} \\
    N_{i,p}(\xi) &=& {\dfrac{\xi - \xi_i}{\xi_{i+p} - \xi_i}  N_{i,p-1}(\xi)} +{\dfrac{\xi_{i+p+1} - \xi}{\xi_{i+p+1} - \xi_{i+1}}  N_{i+1,p-1}(\xi)}. \nonumber
\end{eqnarray}
A B-spline curve can be constructed as the linear combination of the basis functions,

\begin{eqnarray}\label{three}
    C(\xi) &=& \displaystyle\sum_{i=1}^{n}N_{i,p}(\xi)\mathbf{P}_i,
\end{eqnarray}
where $\mathbf{P}_{i}\in\mathbb{R}^{d}$ are the control points associated to the basis functions.

\begin{figure}
    \begin{minipage}{.49\textwidth}
    \includegraphics[scale=0.23]{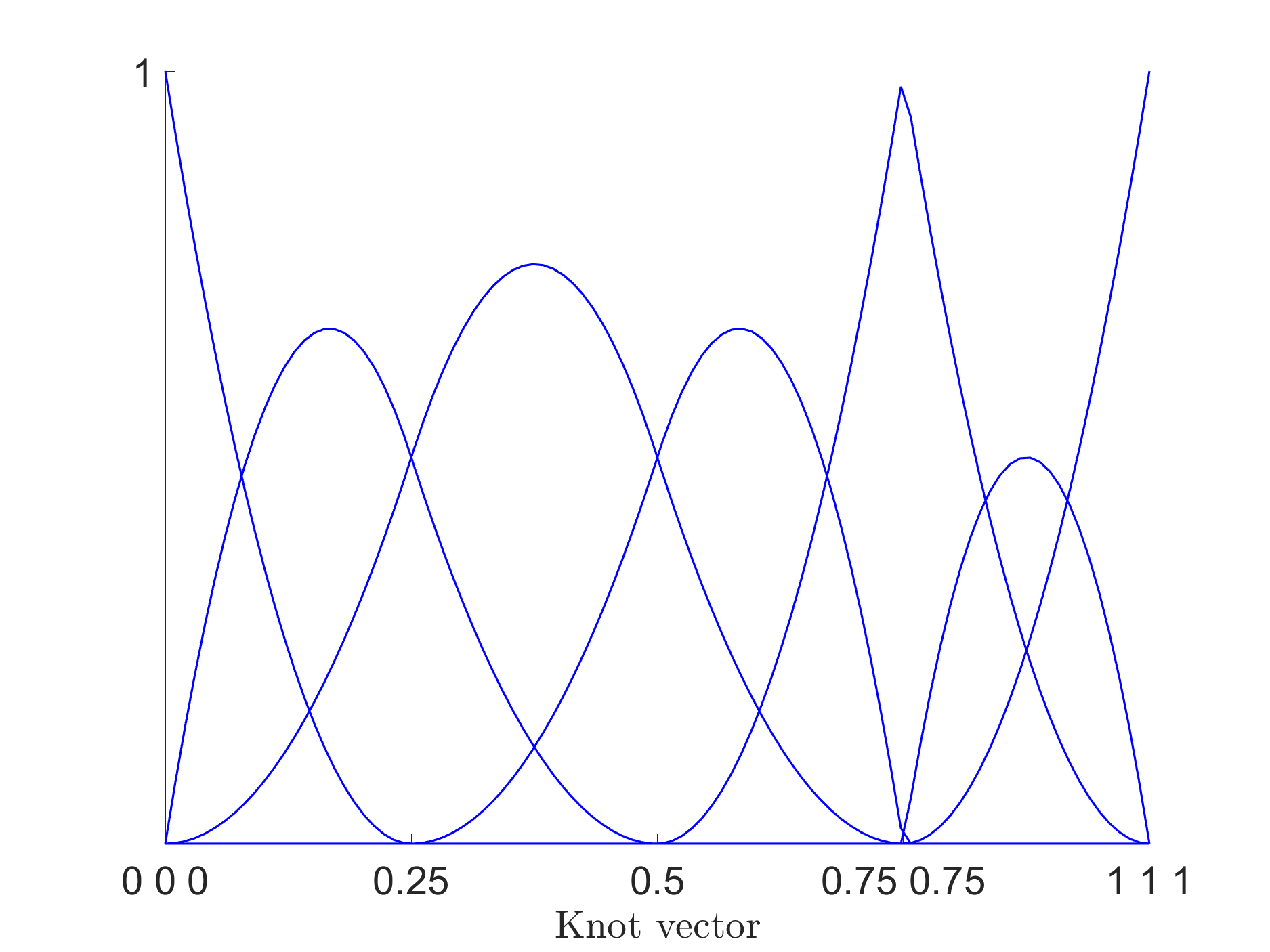}
    \end{minipage}
    \hfil
    \begin{minipage}{.49\textwidth}
    \includegraphics[scale=0.23]{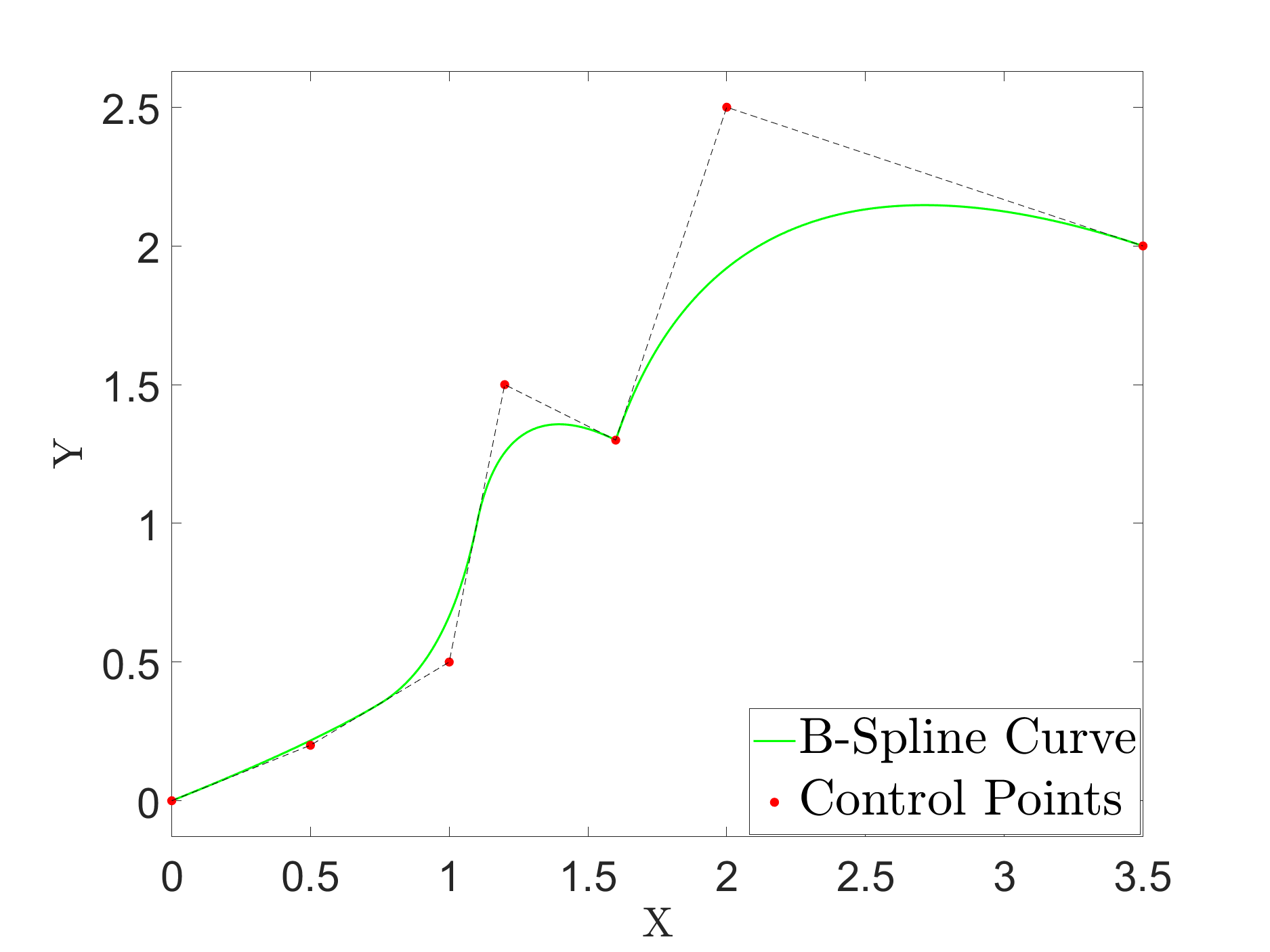}
    \end{minipage}
    \caption{B-spline function with knot vector, $\Xi = [0,0,0,0.25,0.5,0.75,0.75,1,1,1]$ and degree $p=2$. Left: basis functions. Right: curve with control points.}
    \label{fig:example_bspline}
\end{figure}

In general, B-spline functions are piece-wise polynomials, they are non-negative over their domain, they constitute a partition of unity, they have local support, and finally, their continuity depends on knot multiplicity $k$, $C^{p-k}$. In Fig.~\ref{fig:example_bspline}, we have a quadratic B-spline that represent all the aforementioned properties.

For higher dimensions, we need to construct multivariate B-splines as a tensor product of univariate B-splines, as indicated in Fig.~\ref{fig:tensor}.  Considering $\mathbb{R}^{2}$, the degrees $p$ and $q$ and knot vectors $\Xi = \{ \xi_1\leq...\leq \xi_{n+p+1}\}$ and $H = \{ \eta_1\leq...\leq \eta_{m+q+1}\}$, define each dimension of a surface. Thus, two sets of basis functions along with the control net ${\mathbf{P}_{i,j}}$ with $ i=1,2,...,n$, $j=1,2,...,m,$ describe the surface,

\begin{eqnarray}\label{four}
    S(\xi,\eta) &=& \displaystyle\sum_{i=1}^{n}\displaystyle\sum_{j=1}^{m}N_{i,p}(\xi)M_{j,q}(\eta)\mathbf{P}_{i,j}.
\end{eqnarray}
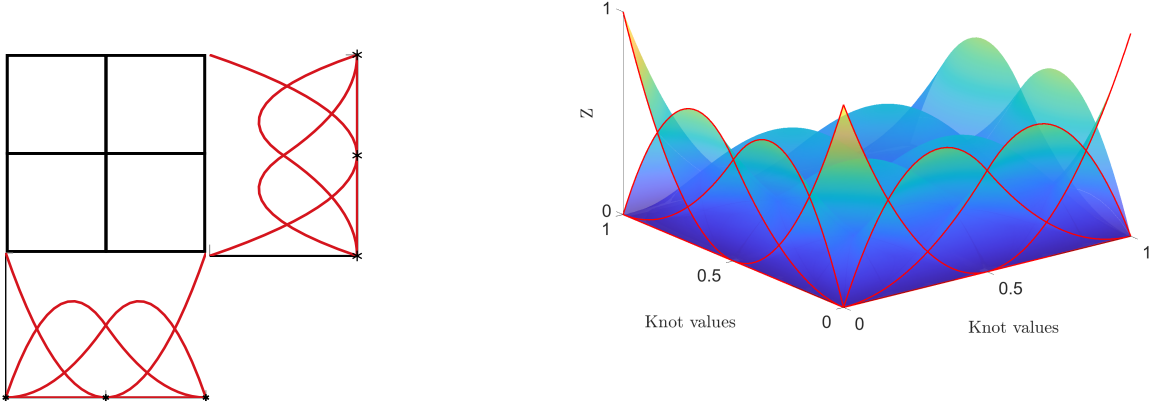
\begin{figure}
\vspace{-1cm}
\begin{minipage}{.43\textwidth}
\begin{center}  
\begin{tikzpicture}[thick,scale=0.65, every node/.style={scale=0.5}]
    \begin{scope}
%
%
\definecolor{mycolor1}{rgb}{0.83500,0.07800,0.11400}%
\definecolor{mycolor2}{rgb}{0.00000,0.44700,0.74100}%
\begin{tikzpicture}

\begin{axis}[%
width=7.641in,
height=10.495in,
at={(1.76in,1.031in)},
scale only axis,
xmin=-1,
xmax=0,
xtick={-3,-2,-1},
ymin=0,
ymax=1,
ytick={1},
axis background/.style={fill=white},
axis x line*=bottom,
axis y line*=right,
yticklabel=\empty,
xticklabel=\empty,
scale=0.1
]
\addplot [color=mycolor1, line width=1.0pt, forget plot]
  table[row sep=crcr]{%
-1	0\\
-0.866825208085612	0.0344827586206897\\
-0.743162901307967	0.0689655172413793\\
-0.629013079667063	0.103448275862069\\
-0.524375743162901	0.137931034482759\\
-0.429250891795482	0.172413793103448\\
-0.343638525564804	0.206896551724138\\
-0.267538644470868	0.241379310344828\\
-0.200951248513674	0.275862068965517\\
-0.143876337693222	0.310344827586207\\
-0.0963139120095124	0.344827586206897\\
-0.0582639714625446	0.379310344827586\\
-0.0297265160523186	0.413793103448276\\
-0.0107015457788346	0.448275862068966\\
-0.00118906064209279	0.482758620689655\\
-0	0.517241379310345\\
-0	0.551724137931034\\
-0	0.586206896551724\\
-0	0.620689655172414\\
-0	0.655172413793103\\
-0	0.689655172413793\\
-0	0.724137931034483\\
-0	0.758620689655172\\
-0	0.793103448275862\\
-0	0.827586206896552\\
-0	0.862068965517241\\
-0	0.896551724137931\\
-0	0.931034482758621\\
-0	0.96551724137931\\
-0	1\\
};
\addplot [color=mycolor1, line width=1.0pt, forget plot]
  table[row sep=crcr]{%
-0	0\\
-0.130796670630202	0.0344827586206897\\
-0.247324613555291	0.0689655172413793\\
-0.349583828775268	0.103448275862069\\
-0.437574316290131	0.137931034482759\\
-0.511296076099881	0.172413793103448\\
-0.570749108204518	0.206896551724138\\
-0.615933412604043	0.241379310344828\\
-0.646848989298454	0.275862068965517\\
-0.663495838287753	0.310344827586207\\
-0.665873959571938	0.344827586206897\\
-0.653983353151011	0.379310344827586\\
-0.62782401902497	0.413793103448276\\
-0.587395957193817	0.448275862068966\\
-0.532699167657551	0.482758620689655\\
-0.466111771700357	0.517241379310345\\
-0.401902497027348	0.551724137931034\\
-0.342449464922711	0.586206896551724\\
-0.287752675386445	0.620689655172414\\
-0.237812128418549	0.655172413793103\\
-0.192627824019025	0.689655172413793\\
-0.152199762187871	0.724137931034483\\
-0.116527942925089	0.758620689655172\\
-0.0856123662306778	0.793103448275862\\
-0.0594530321046374	0.827586206896552\\
-0.0380499405469679	0.862068965517241\\
-0.0214030915576695	0.896551724137931\\
-0.00951248513674208	0.931034482758621\\
-0.00237812128418557	0.96551724137931\\
-0	1\\
};
\addplot [color=mycolor1, line width=1.0pt, forget plot]
  table[row sep=crcr]{%
-0	0\\
-0.00237812128418557	0.0344827586206897\\
-0.00951248513674208	0.0689655172413793\\
-0.0214030915576695	0.103448275862069\\
-0.0380499405469679	0.137931034482759\\
-0.0594530321046374	0.172413793103448\\
-0.0856123662306778	0.206896551724138\\
-0.116527942925089	0.241379310344828\\
-0.152199762187871	0.275862068965517\\
-0.192627824019025	0.310344827586207\\
-0.237812128418549	0.344827586206897\\
-0.287752675386445	0.379310344827586\\
-0.342449464922711	0.413793103448276\\
-0.401902497027348	0.448275862068966\\
-0.466111771700357	0.482758620689655\\
-0.532699167657551	0.517241379310345\\
-0.587395957193817	0.551724137931034\\
-0.62782401902497	0.586206896551724\\
-0.653983353151011	0.620689655172414\\
-0.665873959571938	0.655172413793103\\
-0.663495838287753	0.689655172413793\\
-0.646848989298454	0.724137931034483\\
-0.615933412604043	0.758620689655172\\
-0.570749108204518	0.793103448275862\\
-0.511296076099881	0.827586206896552\\
-0.437574316290131	0.862068965517241\\
-0.349583828775268	0.896551724137931\\
-0.247324613555291	0.931034482758621\\
-0.130796670630202	0.96551724137931\\
-0	1\\
};
\addplot [color=mycolor1, line width=1.0pt, forget plot]
  table[row sep=crcr]{%
-0	0\\
-0	0.0344827586206897\\
-0	0.0689655172413793\\
-0	0.103448275862069\\
-0	0.137931034482759\\
-0	0.172413793103448\\
-0	0.206896551724138\\
-0	0.241379310344828\\
-0	0.275862068965517\\
-0	0.310344827586207\\
-0	0.344827586206897\\
-0	0.379310344827586\\
-0	0.413793103448276\\
-0	0.448275862068966\\
-0	0.482758620689655\\
-0.00118906064209279	0.517241379310345\\
-0.0107015457788346	0.551724137931034\\
-0.0297265160523186	0.586206896551724\\
-0.0582639714625446	0.620689655172414\\
-0.0963139120095124	0.655172413793103\\
-0.143876337693222	0.689655172413793\\
-0.200951248513674	0.724137931034483\\
-0.267538644470868	0.758620689655172\\
-0.343638525564804	0.793103448275862\\
-0.429250891795482	0.827586206896552\\
-0.524375743162901	0.862068965517241\\
-0.629013079667063	0.896551724137931\\
-0.743162901307967	0.931034482758621\\
-0.866825208085612	0.96551724137931\\
-1	1\\
};
\addplot [color=black, line width=.5pt, only marks, mark size=2.0pt, mark=asterisk, mark options={solid, black}, forget plot]
  table[row sep=crcr]{%
0	0\\
};
\addplot [color=black, line width=.5pt, only marks, mark size=2.0pt, mark=asterisk, mark options={solid, black}, forget plot]
  table[row sep=crcr]{%
0	0.5\\
};
\addplot [color=black, line width=.5pt, only marks, mark size=2.0pt, mark=asterisk, mark options={solid, black}, forget plot]
  table[row sep=crcr]{%
0	1\\
};
\end{axis}

\end{tikzpicture}%
    \end{scope}
     \begin{scope}[shift={(-4,0.3)}]
\draw[line width=1pt] (0,0) rectangle (4,4);

\draw[line width=1pt] (0,2) -- (4,2);

\draw[line width=1pt] (2,0) -- (2,4);

\draw[line width=1pt] (0,0) rectangle (2,2);
\draw[line width=1pt] (2,0) rectangle (4,2);
\draw[line width=1pt] (0,2) rectangle (2,4);
\draw[line width=1pt] (2,2) rectangle (4,4); 
    \end{scope}
    \begin{scope}[shift={(-8.5,-5.3)}]
%
%


\definecolor{mycolor1}{rgb}{0.83500,0.07800,0.11400}%

\begin{axis}[%
width=10.495in,
height=7.641in,
at={(1.76in,1.031in)},
scale only axis,
xmin=0,
xmax=1,
xtick={0,0.5,1},
ymin=0,
ymax=1,
ytick={1},
axis background/.style={fill=white},
axis x line*=bottom,
axis y line*=left,
yticklabel=\empty,
xticklabel=\empty,
scale= 0.152
]

\addplot [color=black, forget plot]
  table[row sep=crcr]{%
2.4999446956997e-13	0\\
0.0263157894739079	0\\
0.0526315789475658	0\\
0.0789473684212237	0\\
0.105263157894882	0\\
0.131578947368539	0\\
0.157894736842197	0\\
0.184210526315855	0\\
0.210526315789513	0\\
0.236842105263171	0\\
0.263157894736829	0\\
0.289473684210487	0\\
0.315789473684145	0\\
0.342105263157803	0\\
0.368421052631461	0\\
0.394736842105119	0\\
0.421052631578776	0\\
0.447368421052434	0\\
0.473684210526092	0\\
0.49999999999975	0\\
};
\addplot [color=black, forget plot]
  table[row sep=crcr]{%
0.50000000000025	0\\
0.526315789473908	0\\
0.552631578947566	0\\
0.578947368421224	0\\
0.605263157894882	0\\
0.631578947368539	0\\
0.657894736842197	0\\
0.684210526315855	0\\
0.710526315789513	0\\
0.736842105263171	0\\
0.763157894736829	0\\
0.789473684210487	0\\
0.815789473684145	0\\
0.842105263157803	0\\
0.868421052631461	0\\
0.894736842105119	0\\
0.921052631578776	0\\
0.947368421052434	0\\
0.973684210526092	0\\
0.99999999999975	0\\
};
\addplot [color=mycolor1, line width=1.5pt, forget plot]
  table[row sep=crcr]{%
0	1\\
0.0344827586206897	0.866825208085612\\
0.0689655172413793	0.743162901307967\\
0.103448275862069	0.629013079667063\\
0.137931034482759	0.524375743162901\\
0.172413793103448	0.429250891795482\\
0.206896551724138	0.343638525564804\\
0.241379310344828	0.267538644470868\\
0.275862068965517	0.200951248513674\\
0.310344827586207	0.143876337693222\\
0.344827586206897	0.0963139120095124\\
0.379310344827586	0.0582639714625446\\
0.413793103448276	0.0297265160523186\\
0.448275862068966	0.0107015457788346\\
0.482758620689655	0.00118906064209279\\
0.517241379310345	0\\
0.551724137931034	0\\
0.586206896551724	0\\
0.620689655172414	0\\
0.655172413793103	0\\
0.689655172413793	0\\
0.724137931034483	0\\
0.758620689655172	0\\
0.793103448275862	0\\
0.827586206896552	0\\
0.862068965517241	0\\
0.896551724137931	0\\
0.931034482758621	0\\
0.96551724137931	0\\
1	0\\
};
\addplot [color=mycolor1, line width=1.5pt, forget plot]
  table[row sep=crcr]{%
0	0\\
0.0344827586206897	0.130796670630202\\
0.0689655172413793	0.247324613555291\\
0.103448275862069	0.349583828775268\\
0.137931034482759	0.437574316290131\\
0.172413793103448	0.511296076099881\\
0.206896551724138	0.570749108204518\\
0.241379310344828	0.615933412604043\\
0.275862068965517	0.646848989298454\\
0.310344827586207	0.663495838287753\\
0.344827586206897	0.665873959571938\\
0.379310344827586	0.653983353151011\\
0.413793103448276	0.62782401902497\\
0.448275862068966	0.587395957193817\\
0.482758620689655	0.532699167657551\\
0.517241379310345	0.466111771700357\\
0.551724137931034	0.401902497027348\\
0.586206896551724	0.342449464922711\\
0.620689655172414	0.287752675386445\\
0.655172413793103	0.237812128418549\\
0.689655172413793	0.192627824019025\\
0.724137931034483	0.152199762187871\\
0.758620689655172	0.116527942925089\\
0.793103448275862	0.0856123662306778\\
0.827586206896552	0.0594530321046374\\
0.862068965517241	0.0380499405469679\\
0.896551724137931	0.0214030915576695\\
0.931034482758621	0.00951248513674208\\
0.96551724137931	0.00237812128418557\\
1	0\\
};
\addplot [color=mycolor1, line width=1.5pt, forget plot]
  table[row sep=crcr]{%
0	0\\
0.0344827586206897	0.00237812128418557\\
0.0689655172413793	0.00951248513674208\\
0.103448275862069	0.0214030915576695\\
0.137931034482759	0.0380499405469679\\
0.172413793103448	0.0594530321046374\\
0.206896551724138	0.0856123662306778\\
0.241379310344828	0.116527942925089\\
0.275862068965517	0.152199762187871\\
0.310344827586207	0.192627824019025\\
0.344827586206897	0.237812128418549\\
0.379310344827586	0.287752675386445\\
0.413793103448276	0.342449464922711\\
0.448275862068966	0.401902497027348\\
0.482758620689655	0.466111771700357\\
0.517241379310345	0.532699167657551\\
0.551724137931034	0.587395957193817\\
0.586206896551724	0.62782401902497\\
0.620689655172414	0.653983353151011\\
0.655172413793103	0.665873959571938\\
0.689655172413793	0.663495838287753\\
0.724137931034483	0.646848989298454\\
0.758620689655172	0.615933412604043\\
0.793103448275862	0.570749108204518\\
0.827586206896552	0.511296076099881\\
0.862068965517241	0.437574316290131\\
0.896551724137931	0.349583828775268\\
0.931034482758621	0.247324613555291\\
0.96551724137931	0.130796670630202\\
1	0\\
};
\addplot [color=mycolor1, line width=1.5pt, forget plot]
  table[row sep=crcr]{%
0	0\\
0.0344827586206897	0\\
0.0689655172413793	0\\
0.103448275862069	0\\
0.137931034482759	0\\
0.172413793103448	0\\
0.206896551724138	0\\
0.241379310344828	0\\
0.275862068965517	0\\
0.310344827586207	0\\
0.344827586206897	0\\
0.379310344827586	0\\
0.413793103448276	0\\
0.448275862068966	0\\
0.482758620689655	0\\
0.517241379310345	0.00118906064209279\\
0.551724137931034	0.0107015457788346\\
0.586206896551724	0.0297265160523186\\
0.620689655172414	0.0582639714625446\\
0.655172413793103	0.0963139120095124\\
0.689655172413793	0.143876337693222\\
0.724137931034483	0.200951248513674\\
0.758620689655172	0.267538644470868\\
0.793103448275862	0.343638525564804\\
0.827586206896552	0.429250891795482\\
0.862068965517241	0.524375743162901\\
0.896551724137931	0.629013079667063\\
0.931034482758621	0.743162901307967\\
0.96551724137931	0.866825208085612\\
1	1\\
};
\addplot [color=black, line width=1.0pt, only marks, mark size=2.0pt, mark=asterisk, mark options={solid, black}, forget plot]
  table[row sep=crcr]{%
0	0\\
};
\addplot [color=black, line width=1.0pt, only marks, mark size=2.0pt, mark=asterisk, mark options={solid, black}, forget plot]
  table[row sep=crcr]{%
0.5	0\\
};
\addplot [color=black, line width=1.0pt, only marks, mark size=2.0pt, mark=asterisk, mark options={solid, black}, forget plot]
  table[row sep=crcr]{%
1	0\\
};
\end{axis}

    \end{scope}
\end{tikzpicture}
\end{center}
\end{minipage}
\hfil
\begin{minipage}{.55\textwidth}
    \includegraphics[scale=0.25]{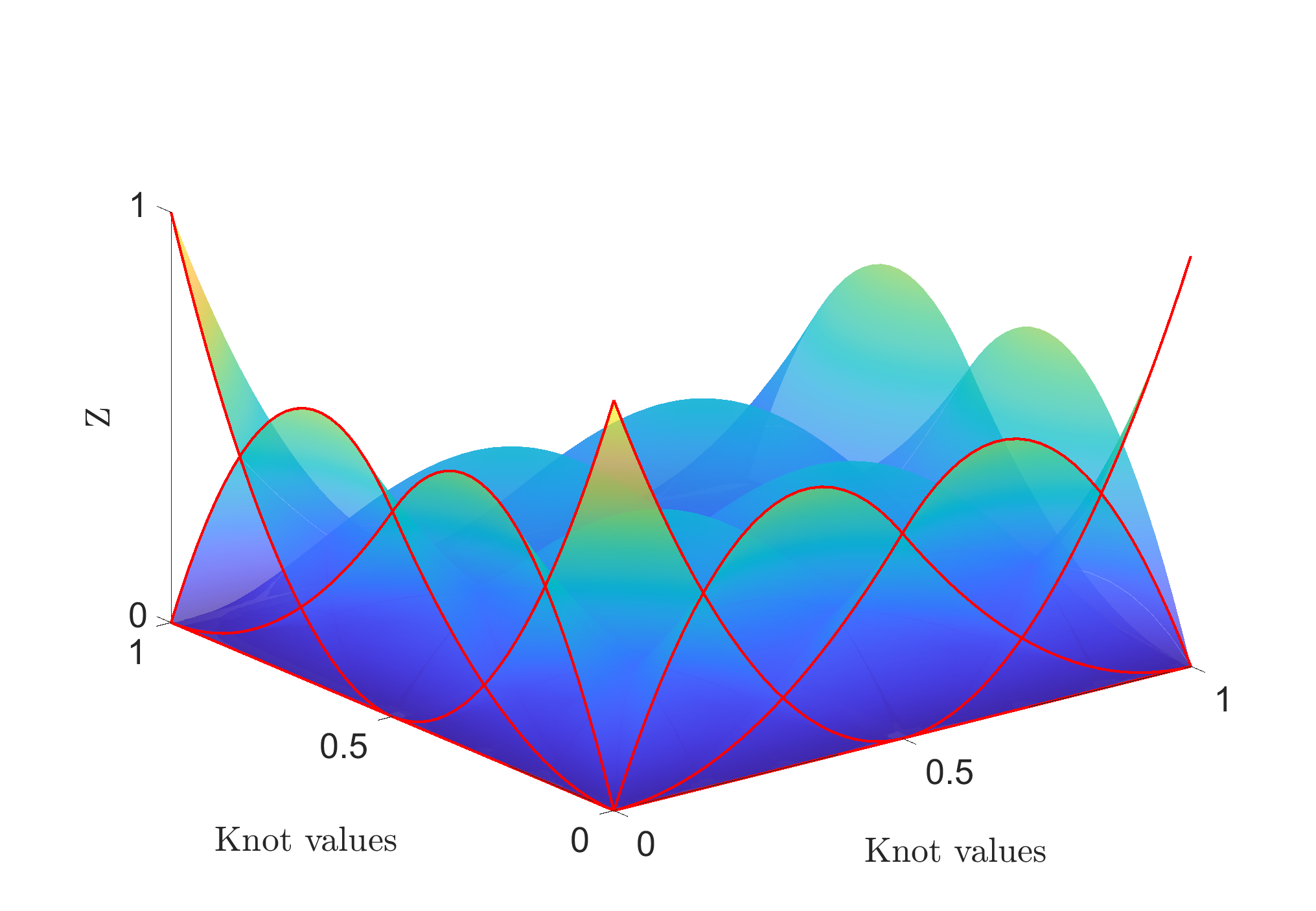}
\end{minipage}
\caption{Reference domain with the basis function in 2D, as tensor product structure.}
\label{fig:tensor}
\end{figure}

\subsection{Multi-level structure with THB-splines}

\begin{figure}[h]
    \begin{minipage}{.49\textwidth}
    \includegraphics[scale=0.25]{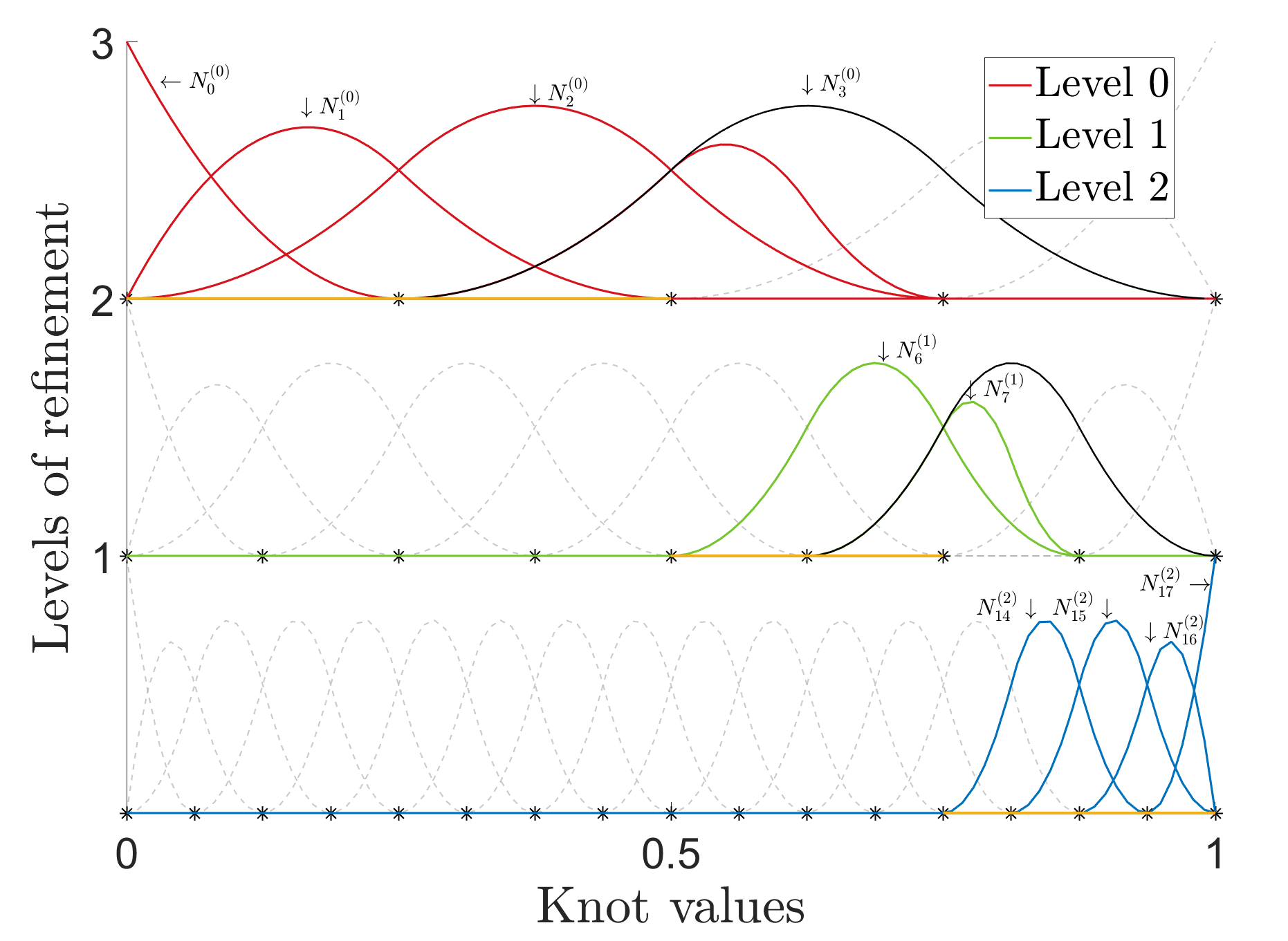}
    \end{minipage}
    \hfil
    \begin{minipage}{.49\textwidth}
    \includegraphics[scale=0.28]{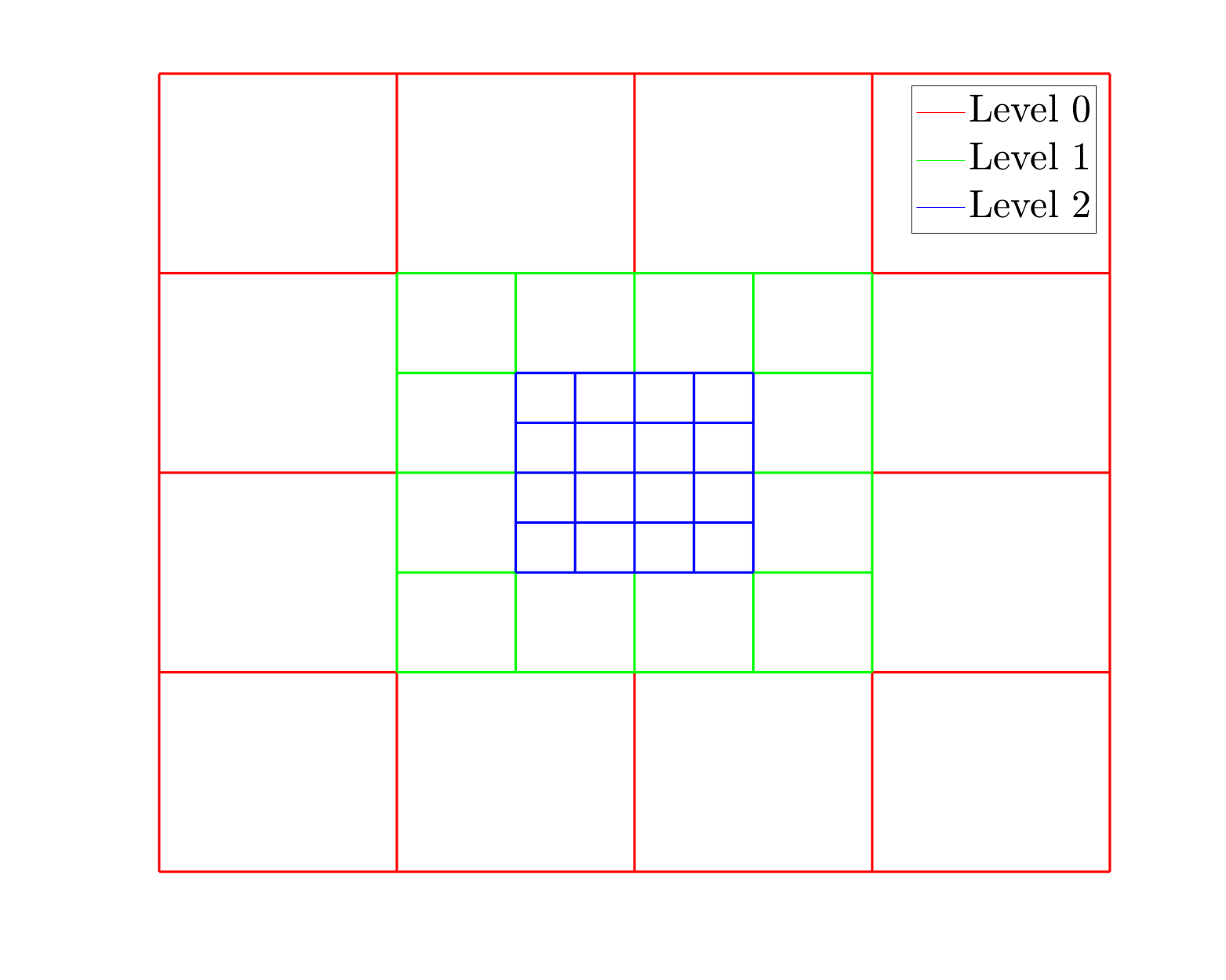}
    \end{minipage}
    \caption{Left: a univariate quadratic THB-spline with two additional levels of refinement. Right: a 2D representation of a tensor product quadratic THB-splines.}
    \label{fig:THB}
\end{figure}

A multi-level structure with hierarchical B-splines offers a state-of-the-art method for adaptive refinement, especially in representing complex geometries. Knot insertion algorithms construct the individual levels, $\ell$. We follow the dyadic refinement technique with $2d \cdot n_{el}$ elements for each new level of refinement ($d$ is the dimension and $n_{el}$ number of elements in $\ell-1$), so the levels preserve the same polynomial degree. This multi-level structure consists of sequence of nested spline spaces of univariate splines, $\mathbb{S}^{(0)}(\Omega) \subset \mathbb{S}^{(1)}(\Omega) \dots \subset  \mathbb{S}^{(\ell_{max})}(\Omega)$. In Fig.~\ref{fig:THB}, a univariate quadratic B-spline with two additional levels of refinement is depicted. The geometry and the parametric representation do not change under the refinement process, but the relation between the levels occurs by an operator called subdivision matrix $\mathbf{S}$. The parent functions, $N(\xi)_{i,p}^{(\ell)}$ of the level $\ell$ are a linear combination of fine-scale functions $\hat{N}_{j,p}(\xi)^{(\ell+1)}$ (children functions) of level $\ell + 1$,

\begin{eqnarray}\label{five}
        N(\xi)_{i,p}^{(\ell)} &=& \displaystyle\sum_{j=1}^{\hat{n}} \mathbf{S_{j,i}} \hat{N}_{j,p}(\xi)^{(\ell+1)}. 
\end{eqnarray}

THB-splines \cite{THB} enhance traditional hierarchical B-splines by restoring the partitioned unity property and thereby reducing their overlap. As a result, THB-splines provide a more flexible and powerful tool for local refinement. The black basis functions in Fig.~\ref{fig:THB}, are the overlapping ones, where we apply the truncation mechanism.  The truncated basis functions are defined as,

\begin{eqnarray}\label{six}
    trunc N(\xi)_{i,p}^{(\ell)} &=& \displaystyle\sum_{j \in \eta^{\ell+1}_D} S^{(\ell)}_{ji} N(\xi)_{j,p}^{(\ell+1)}.
\end{eqnarray}

As previously stated, in a hierarchical structure, each basis function at level $\ell$ is a linear combination of basis functions at level $\ell+1$, encoded in the subdivision matrix \eqref{five}. At that point, the global multi-level extraction operator, $\mathbf{M}^{glob}_{\ell}$, can be defined by applying a sequence of nested spline spaces, where the goal is to extract the hierarchical functions supporting each element at the specific level. For levels $\ell=0,\dots,\ell_{max}$ the $\mathbf{M}^{glob}_{\ell_{max}}$ can be obtained by joining the rows, corresponding to the active basis, of the refinement operators $\mathbf{S}^{(\ell,\ell_{max})}$ \cite{MBE1}. For example, the multi-level operator for level 2, $\mathbf{M}^{(\ell_2)}$, is derived by the operators $\mathbf{S}^{(\ell_0,\ell_1)},\mathbf{S}^{(\ell_1,\ell_2)},\mathbf{S}^{(\ell_2,\ell_2)}$ and $\mathbf{M}^{(\ell_2)} \in \mathbb{R}^{n\times m}$, where $n$ is the active basis functions of $\ell_0$ plus the the active basis functions of $\ell_1$ and $\ell_2$ and $m$ is the total number of basis functions of $\ell_2$.

\subsection{Standard B\'{e}zier extraction}

Similar to standard finite elements, there is a desire for an element-by-element assembly routine to increase the computational efficiency of the assembly procedure. One way to achieve this is B\'{e}zier extraction, which maps a piecewise Bernstein polynomial basis into a B-spline basis and vice versa \cite{BE}. This mapping allows using $C^0$ B\'{e}zier elements as the finite element representation of smooth splines.

\begin{figure}
    \begin{minipage}{.4\textwidth}
                \includegraphics[scale = 0.2]{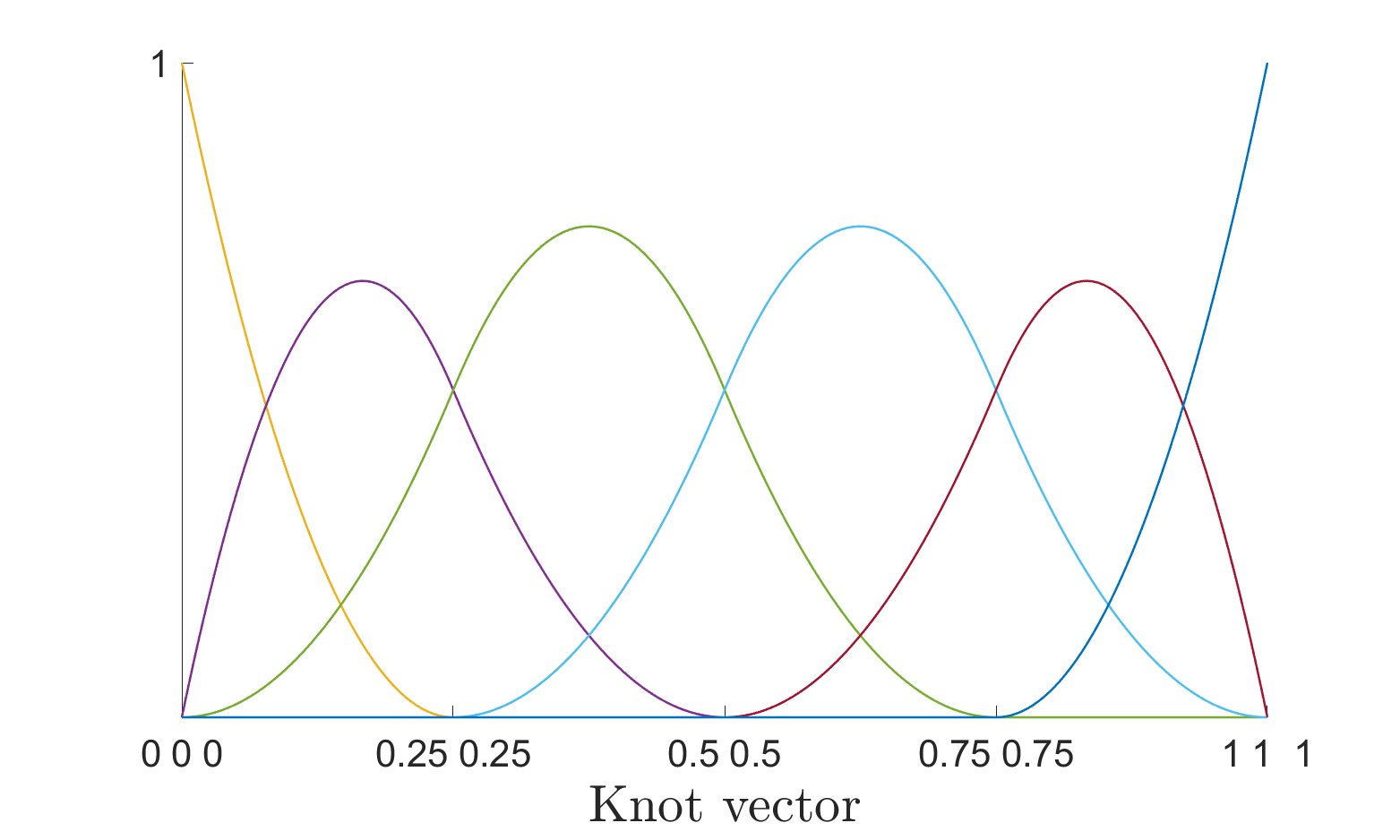}\\
                \includegraphics[scale = 0.2]{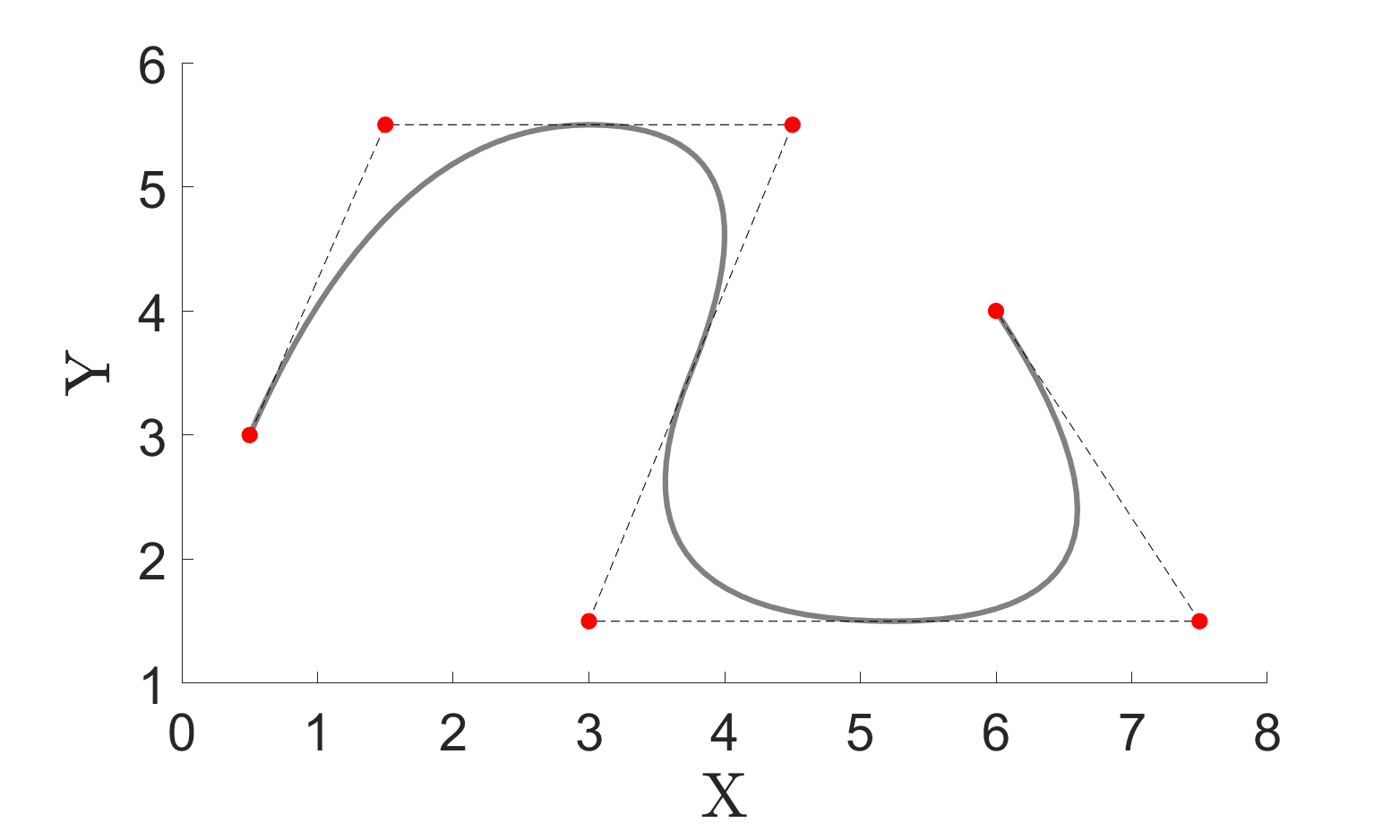}
            \end{minipage}
            \hfill
            \begin{minipage}{.15\textwidth}
                \begin{center}
                    $\mathbf{E}$\\
                    \begin{tikzpicture}[>=latex]
                        
                        \draw[red,->] (0,0) arc
                        [
                            start angle=160,
                            end angle=20,
                            x radius=0.8cm,
                            y radius =0.5cm
                        ] ;
                        \draw[red,<-] (0,0) arc
                        [
                            start angle=-160,
                            end angle=-20,
                            x radius=0.8cm,
                            y radius =0.5cm
                        ] ;

                    \end{tikzpicture}\\
                    \vspace{-0.1cm}
                        $\mathbf{E^\intercal}$ \\
                        $\mathbf{N = EB}$
                    
                \end{center}   
            \end{minipage}
            \hfill
            \begin{minipage}{.4\textwidth}
                \includegraphics[scale = 0.2]{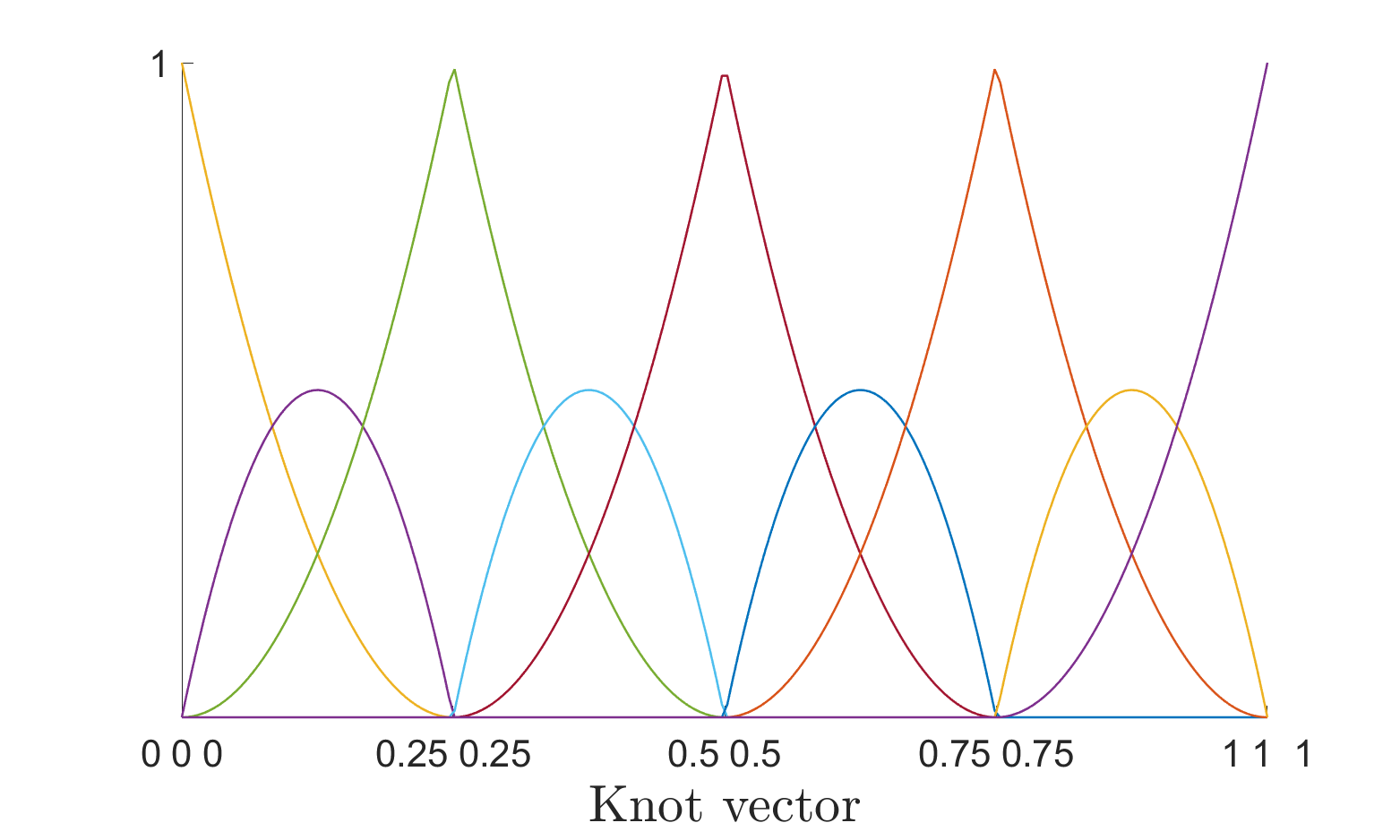}\\
                \includegraphics[scale = 0.2]{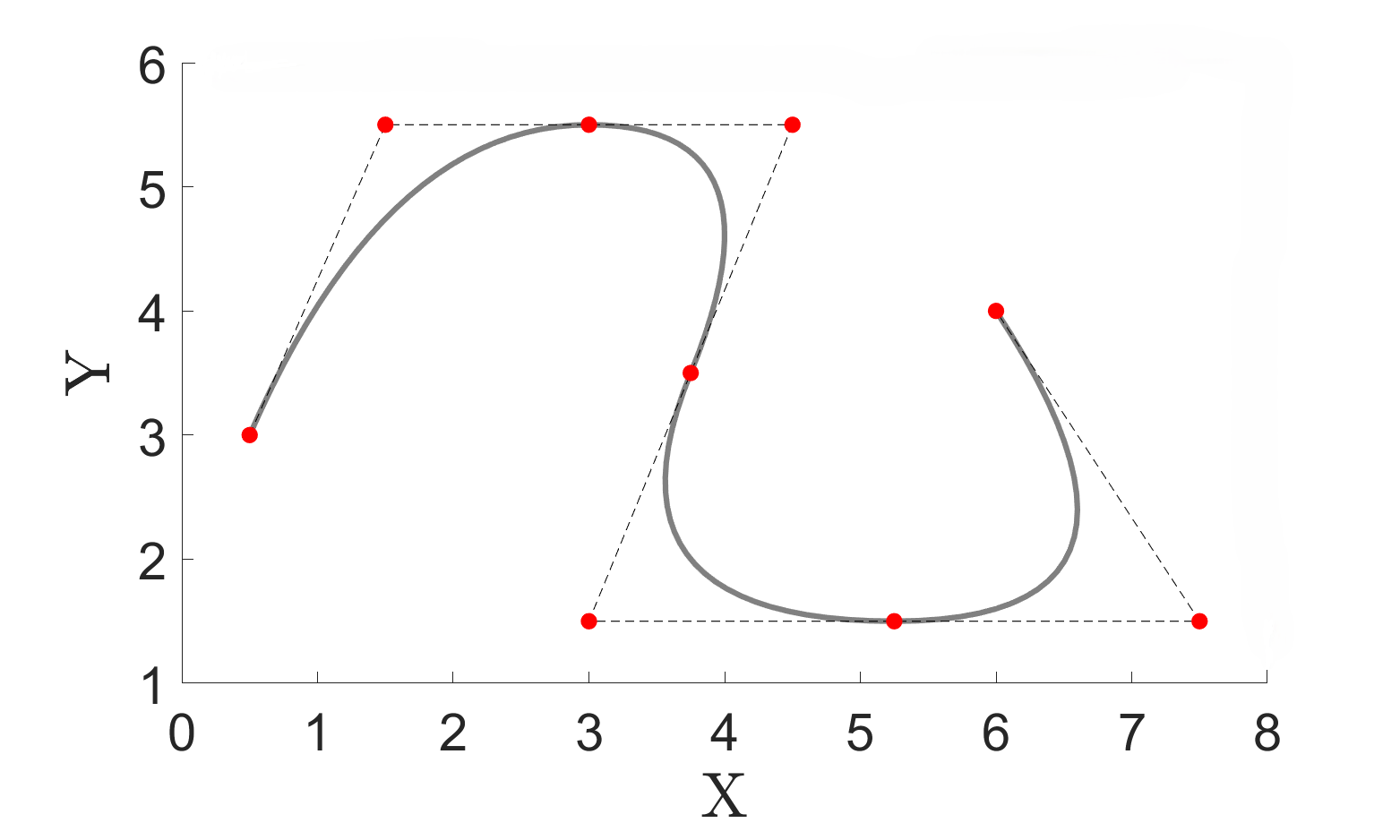}
            \end{minipage}
            \caption{Left: B-spline basis and curve. Right: B\'{e}zier basis and curve.}
            \label{fig:MAP}
\end{figure}
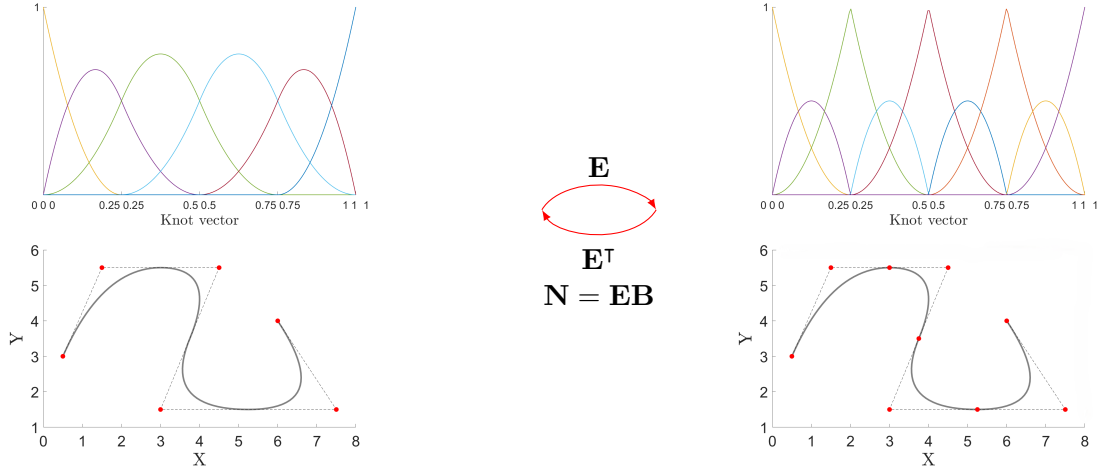

We use B\'{e}zier decomposition, which is accomplished by repeating all interior knots of a knot vector until they have a multiplicity equal to the degree of the polynomial \cite{BE}. A representation of B\'{e}zier extraction for a univariate B-spline function is depicted in Fig.~\ref{fig:MAP}. An important feature is that the initial curve geometry remains unchanged by changing only the basis and the corresponding control points. The B-spline functions, $\mathbf{N}$, can be expressed as a linear combination of Bernstein polynomials $\mathbf{B}$ with the B\'{e}zier extraction operator $\mathbf{E}$.
    
\section{Multi-level B\'{e}zier extraction}
The combination of those two techniques of multi-level THB-splines and B\'{e}zier extraction results in multi-level B\'{e}zier extraction, which decompose B-spline basis functions into a series of simpler B\'{e}zier elements across multiple hierarchical levels.

To construct the multi-level B\'{e}zier extraction operator, $\mathbf{C}$, for the spline basis, we localize the global multi-level extraction operator $\mathbf{M}^{loc}_{e}$ and the B\'{e}zier extraction operator $\mathbf{E}^{loc}_{e}$, for an element-by-element approach. This creates a direct map $\mathbf{C}^{e}$ from a standard set of reference basis functions equal for each element (Bernstein basis, $\mathbf{B}$) to the multi-level local basis (hierarchical basis, $\mathbf{H^e}$),
\begin{eqnarray}\label{seven}
    \mathbf{H^e} = \mathbf{M}^{loc}_{e} \mathbf{E}^{loc}_{e} \mathbf{B} = \mathbf{C}^{e} \mathbf{B}.
\end{eqnarray}

\begin{figure}
    \begin{minipage}{.33\textwidth}
    \includegraphics[scale=0.25]{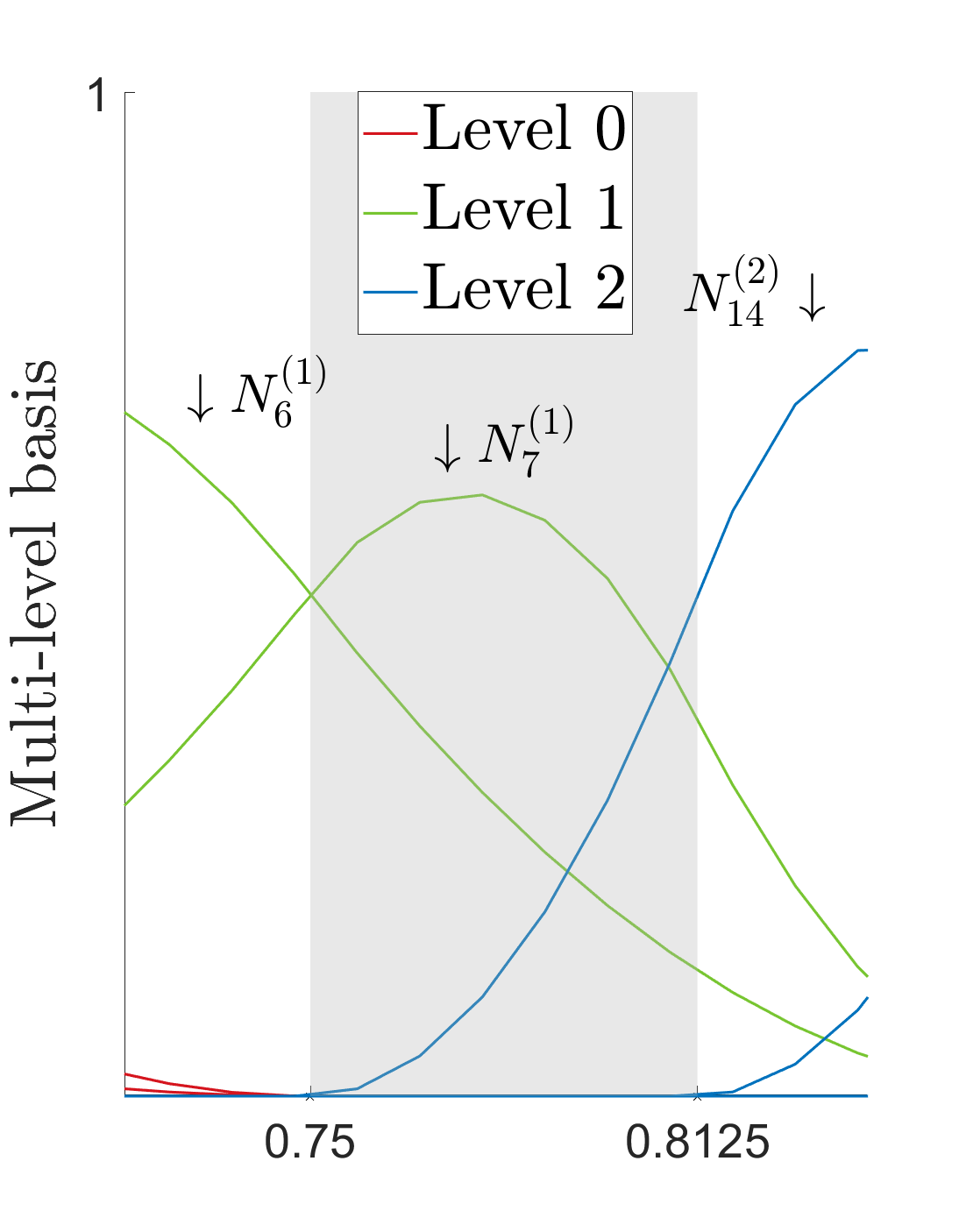}
    \end{minipage}
    \hfil
    \begin{minipage}{.3\textwidth}
    $\begin{bmatrix}
                N_{6}^{(1)} \\
                N_{7}^{(1)} \\
                N_{14}^{(2)}
            \end{bmatrix} = \mathbf{M}^{loc}_{e_3} \mathbf{E}^{loc}_{e_3}  \begin{bmatrix}
            B_{0}\\
            B_{1}\\
            B_{2} 
        \end{bmatrix}$
    \end{minipage}
    \hfil
    \begin{minipage}{.33\textwidth}
    \includegraphics[scale=0.25]{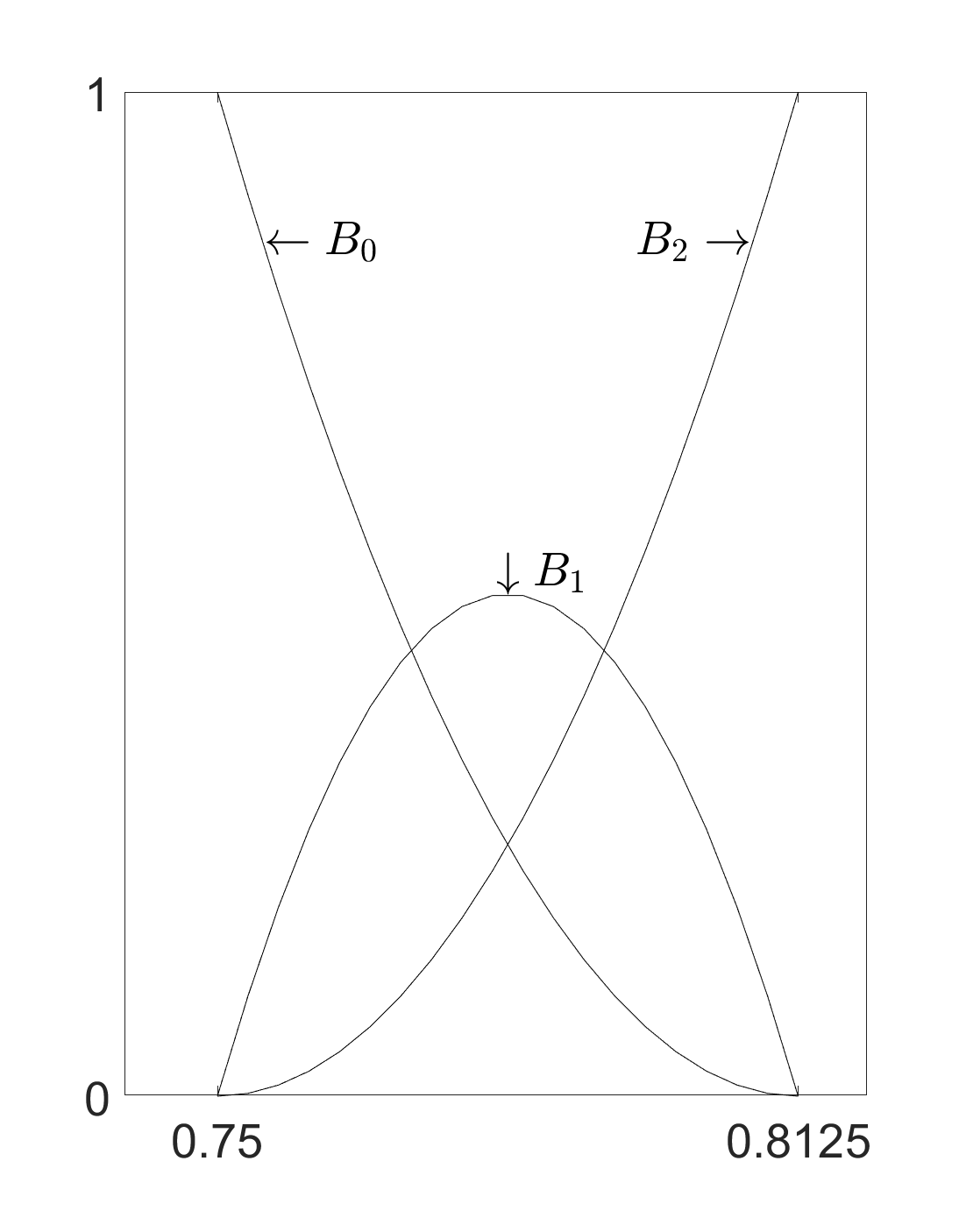}
    \end{minipage}
    \caption{Translation from hierarchical spline space to local Bernstein polynomials.}
    \label{fig:translation}
\end{figure}

In Fig.~\ref{fig:translation}, there is a uni-variate B-spline element on the left side with basis functions from different levels of the refinement. This element can be transformed directly to a B\'{e}zier element with Bernstein basis functions only through the application of the operator $\mathbf{C}^{e_3} = \mathbf{M}^{loc}_{e_3} \mathbf{E}^{loc}_{e_3}$.

The same procedure can be extended into multipatch domains, where the computation domain $\Omega$ is divided into $n_i$ non-overlapping subdomains $\Omega_i$ or patches. Therefore more complex geometries can be handled with greater detail. The geometrical mapping from the reference domain, $\Phi_i$, corresponds to each patch to construct the physical geometry. Section \ref{sec:exam} presents a numerical example of a multipatch domain for a scalar magnetostatic 2D problem.

\subsection{Problem assembly for numerical simulation}
The multi-level B\'{e}zier extraction operator can be used to assemble the system of matrices and vectors of the discretized problem. For instance, the calculation of the stiffness matrix of B-spline elements usually requires special types of algorithms, able to handle elements with a non-standard set of basis and hierarchical structure. In this work, GeoPDEs \cite{GeoPDEs}  is employed due to its implementation of local refinement techniques in IGA with THB-splines. \\
On the other hand, if B\'{e}zier elements are considered, allow for an element-wise assembly as in standard finite element solvers, we can compute a stiffness matrix of a classic B\'{e}zier element and then transform it to a B-spline element stiffness matrix. For this, a local element-by-element approach needs to be taken into account, and then with the proper transformation matrix the global stiffness matrix of the THB-spline basis is computed, 
\begin{eqnarray}\label{eight}
    \mathbf{K}_{B-spline}^e &=& \mathbf{C}^{e}  \mathbf{K}_{Bezier}  (\mathbf{C}^{e})^T \\
    \mathbf{K}_{global} &=& \displaystyle\sum_{e=1}^{n_{el}} \mathcal{T}(\mathbf{K}_{B-spline}^e).\nonumber
\end{eqnarray}
The $\mathbf{K}_{Bezier}$ is the stiffness matrix of a B\'{e}zier element on the corresponding level of refinement. Together with the local multi-level B\'{e}zier extraction operators, $\mathbf{C}^{e}$, the stiffness matrix of the THB-spline space element is calculated, $\mathbf{K}_{B-spline}^e$.

\subsection{Posteriori error estimator}
\label{sec:err}
Let us now look at a posteriori error estimator based on the least-square formulation \cite{est}. Here the refinement follows consecutive refinement levels to consider the areas of interest locally. In brief, first, the initial discretized system needs to be solved,

\begin{eqnarray}\label{eleven} 
    \vspace{ -.1cm} \begin{bmatrix}
        A_h & B_h \\ 
        B^T_h & 0
    \end{bmatrix}
    \begin{bmatrix}
        \underline{p}\\ 
        \underline{u}
    \end{bmatrix} &=&
    \begin{bmatrix}
        \underline{f}\\ 
        0
    \end{bmatrix}.
\end{eqnarray}

Here, $A_h  \in \mathbb{R}^{M_Y \times M_Y},$ is the stiffness matrix of the fine mesh, $B_h \in \mathbb{R}^{M_Y \times M_X},$ is the stiffness matrix of the coarse mesh, mapped onto the fine mesh and $f$ is the source term. The solution vector consists of the error approximation, $p$, and the numerical solution $u$. With the error approximation, we evaluate the local error indicators into the coarse mesh,

\begin{eqnarray}\label{twelve} 
     \vspace{ -.1cm} \eta_k^2 &:=& \int \left | \nabla{p_h(x))} \right |^2 {\mathrm{d} x}.
\end{eqnarray}

Finally, a marking strategy determines the elements that need refinement. In this work, the Dörfler criterion is applied with a minimal set $M \subset N$ where $N$, is all elements in the coarse mesh and $M$, is the subset of elements marked for refinement based on the error indicators,

\begin{eqnarray}\label{twelve} 
     \sum_{k \in M}^{} \eta_k^2 &\geqslant& \theta \sum_{k \in N}^{} \eta_k^2,
\end{eqnarray}
where $\theta$ is a parameter in the range $0<\theta<1$. 

\begin{figure}[h!]
    \begin{minipage}{.43\textwidth}
        \includegraphics[scale=0.23]{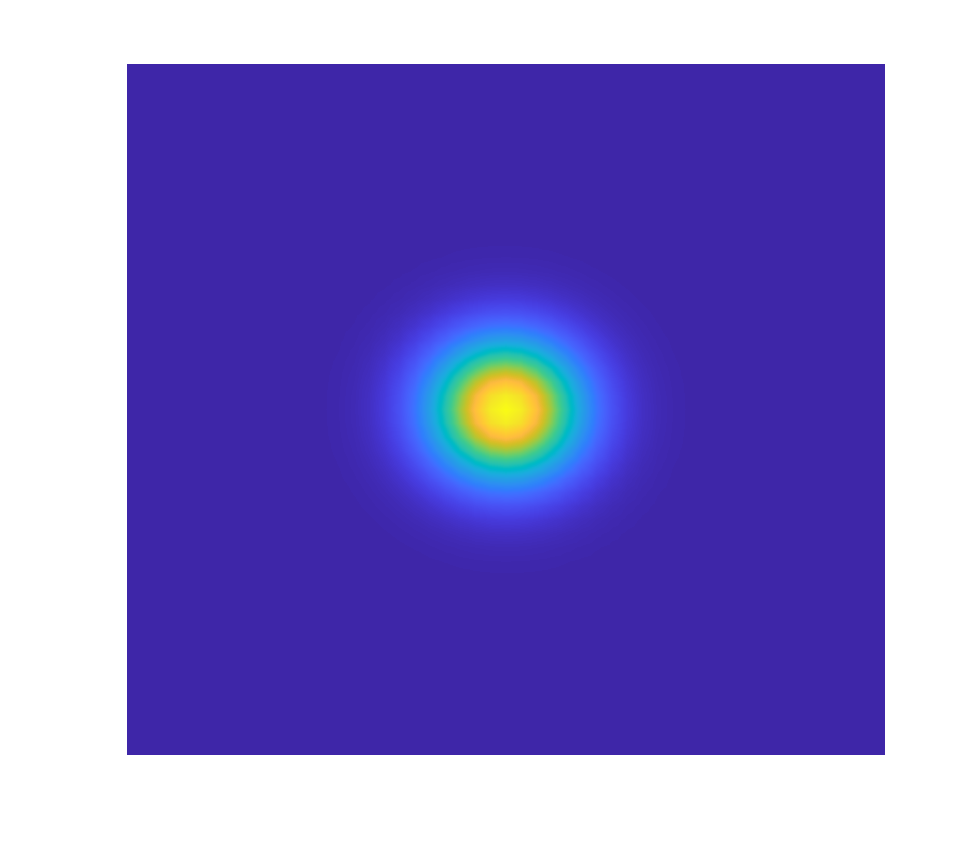} \\  \includegraphics[scale=0.255]{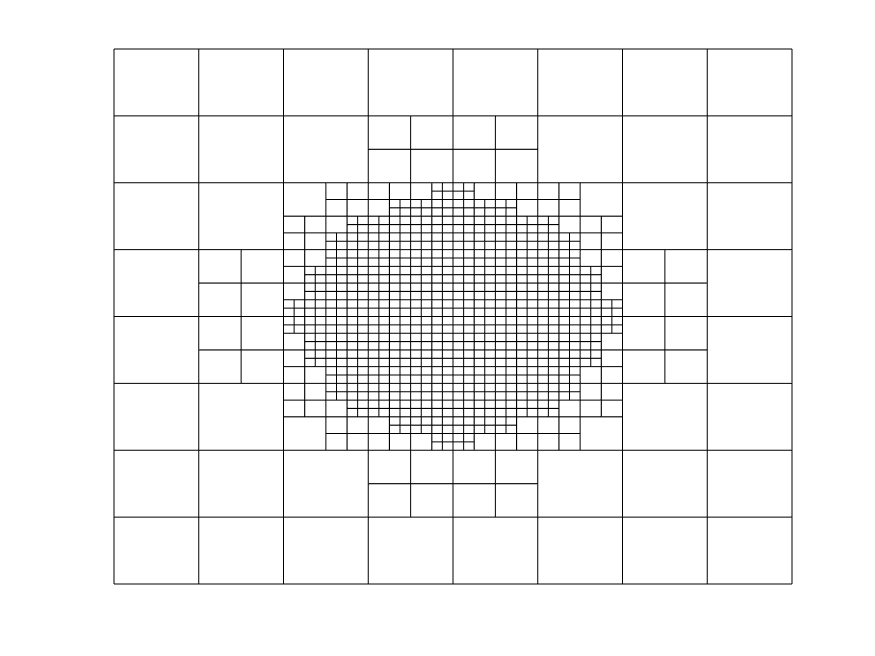}    
    \end{minipage}
    \hfil
    \begin{minipage}{.55\textwidth}
        \includegraphics[scale=0.235]{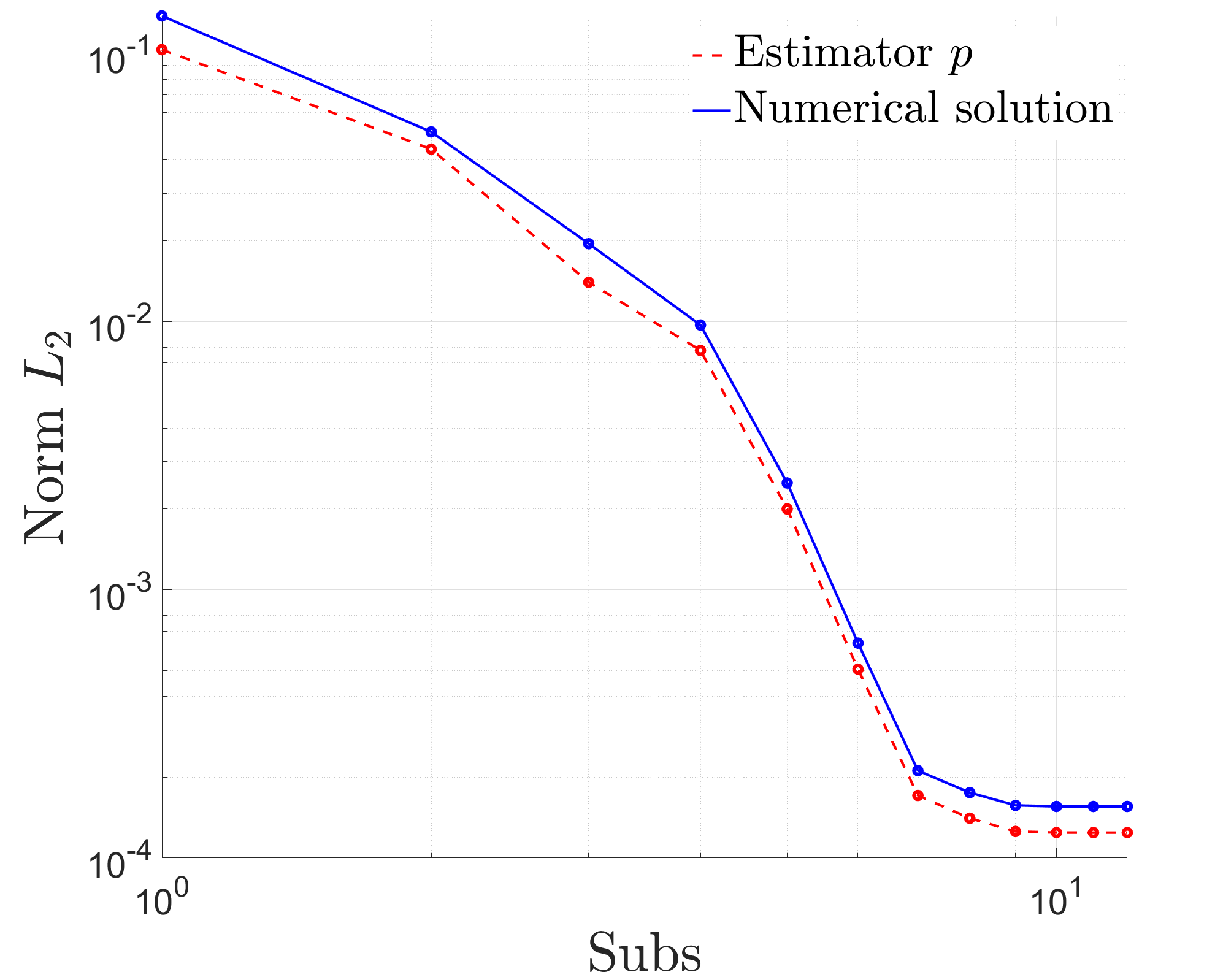}
    \end{minipage}
    \caption{Estimator results for the Poisson problem.}
    \label{fig:err_est}
\end{figure}

As an example of this error estimator, a Poisson problem with non-homogeneous Dirichlet boundary conditions in a unit square domain with a peak value in the middle was computed. The results in Fig.~\ref{fig:err_est}, measure the true error between the numerical and the exact solution (blue line) and the behavior of the estimator (red line). The two lines converge similar and that indicates that the error estimator captures the behavior of the true error in the numerical solution. This gives the potential to extend the use of the error estimator to more complex problems with more detailed geometries.

\section{Numerical example}
\label{sec:exam}
\begin{figure}[h]
    \centering\includegraphics[width=8.3cm,height=6.3cm]{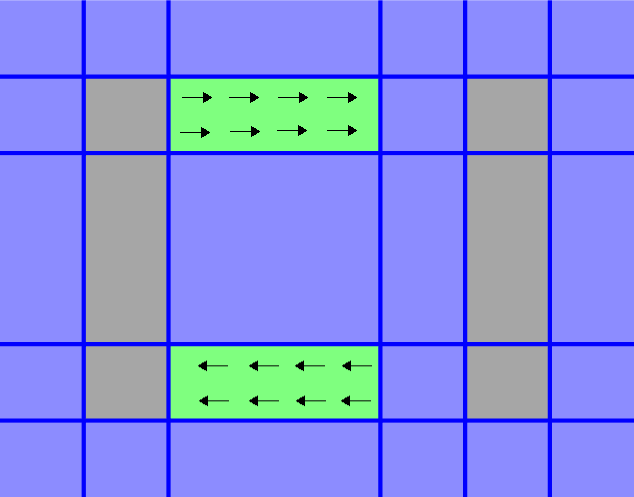}
    \caption{The thick lines indicate the boundaries of the patches. The variation in colors represents the different materials. The blue is the air, the grey is the iron and the light green is the neodymium magnet. The arrows indicate the direction of the magnetic field.}
    \label{fig:geometry}
\end{figure}

GeoPDEs \cite{GeoPDEs}, an open-source FE solver for IGA, implemented in Matlab/Octave, is applied for the numerical example of this contribution. We have a 2D scalar magnetostatic problem with a horseshoe magnet and a metal sheet. This is a multipatch geometry with 30 patches and 3 different materials, air, iron, and neodymium magnet, as shown in Fig.~\ref{fig:geometry}.

To solve the magnetostatic problem in magnetic vector potencial form, the following derivation is applied. We start with the Amp\'{e}re–Maxwell equation, given in terms of the magnetic field strength $\mathbf{H}$, where the displacement currents are disregarded, and the Gauss's law for magnetism, $\mathbf{B}$ is the magnetic flux density,
\begin{eqnarray}\label{eight}
    \nabla \times \mathbf{H}  &=& \mathbf{J}_f \\
    \nabla \cdot \mathbf{B} &=& 0. \nonumber
\end{eqnarray}
Then the definition of the magnetic vector potential and the material law (constitutive equation) are introduced.
\begin{eqnarray}\label{eight}
     \mathbf{B} = \nabla \times \mathbf{A}, \implies  \nabla \cdot \mathbf{B} = \nabla \cdot \left( \nabla \times \mathbf{A} \right), \\
     \mathbf{B} = \mu\mathbf{H} + \mathbf{B_r} \Rightarrow \mathbf{H} = \mu^{-1}\left( \mathbf{B} - \mathbf{B_r} \right), \text{with } \mu = \mu_0\mu_r\nonumber,
\end{eqnarray}
where $\mu_0$ is the vacuum permeability and $\mu_r$ is the relative permeability.
The 3D equations reduce to a 2D form by assuming that all variations along the $z$-axis are zero and the fields vary only over the $xy$-plane. By expressing the magnetic flux density, $\mathbf{B}$, in terms of the vector potential, $\mathbf{A}$, and substituting it into the constitutive relations and Amp\'{e}re–Maxwell Law, we derive a scalar differential equation.
\begin{eqnarray}\label{eight}
    \nabla \times  \left( \mu^{-1} \left( \nabla \times \mathbf{A} - \mathbf{B_r}\right)\right) &=& \mathbf{J},
\end{eqnarray}
and in the 2D case,
\begin{eqnarray}\label{eight2}
    \frac{\partial}{\partial x } \left ( \mu^{-1} \left ( -\frac{\partial A_z }{\partial x} - B_{ry} \right )\right ) - \frac{\partial}{\partial y } \left ( \mu^{-1} \left ( \frac{\partial A_z }{\partial y} - B_{rx} \right )\right ) &=& J_z.
\end{eqnarray}
Note that in 3D edge elements \eqref{eight2} must be used to discretize \eqref{eight} \cite{mag}, in 2D nodal elements are sufficient.

\begin{figure}[h]
    \centering\includegraphics[width=8.3cm,height=6.3cm]{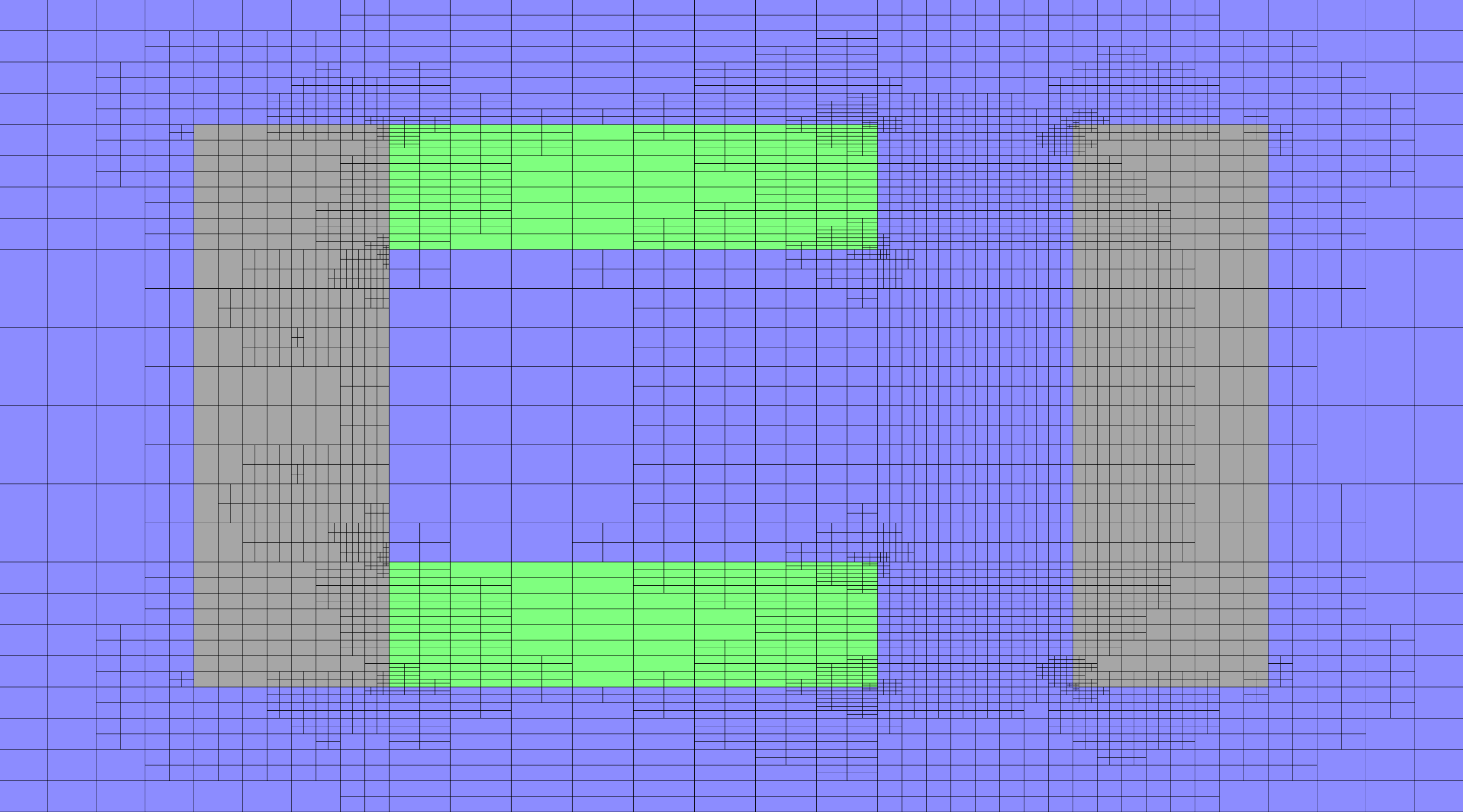}
    \caption{Local mesh refinement.}
    \label{fig:mesh}
\end{figure}

To demonstrate the correctness of the B\'{e}zier approach, we do not apply the error estimator of Section \ref{sec:err} but use a numerical reference solution. The reference solution is obtained using a uniformly refined mesh with a high density of elements and many degrees of freedom to the specific problem (No.~elements = 108000, DoFs = 112302). With that reference solution $u$, the true error approximation with a prescribed tolerance is applied. The quality of the approximate solution $u_h$ is assessed by the $L_2$-error,
\begin{eqnarray}\label{ten}
    \epsilon_{L_2} &=& \left( \int_\Omega |u - u_h|^2 \,dx \right)^{1/2}.
\end{eqnarray}
An element is marked for the refinement if,
\begin{eqnarray}\label{ten}
    \epsilon_{L_2,n} &>& tolerance =  10^{-8}.
\end{eqnarray}
The final locally refined mesh is in Fig.~\ref{fig:mesh}.

\begin{figure}[h]
    \begin{minipage}{.49\textwidth}
        \includegraphics[scale=0.26]{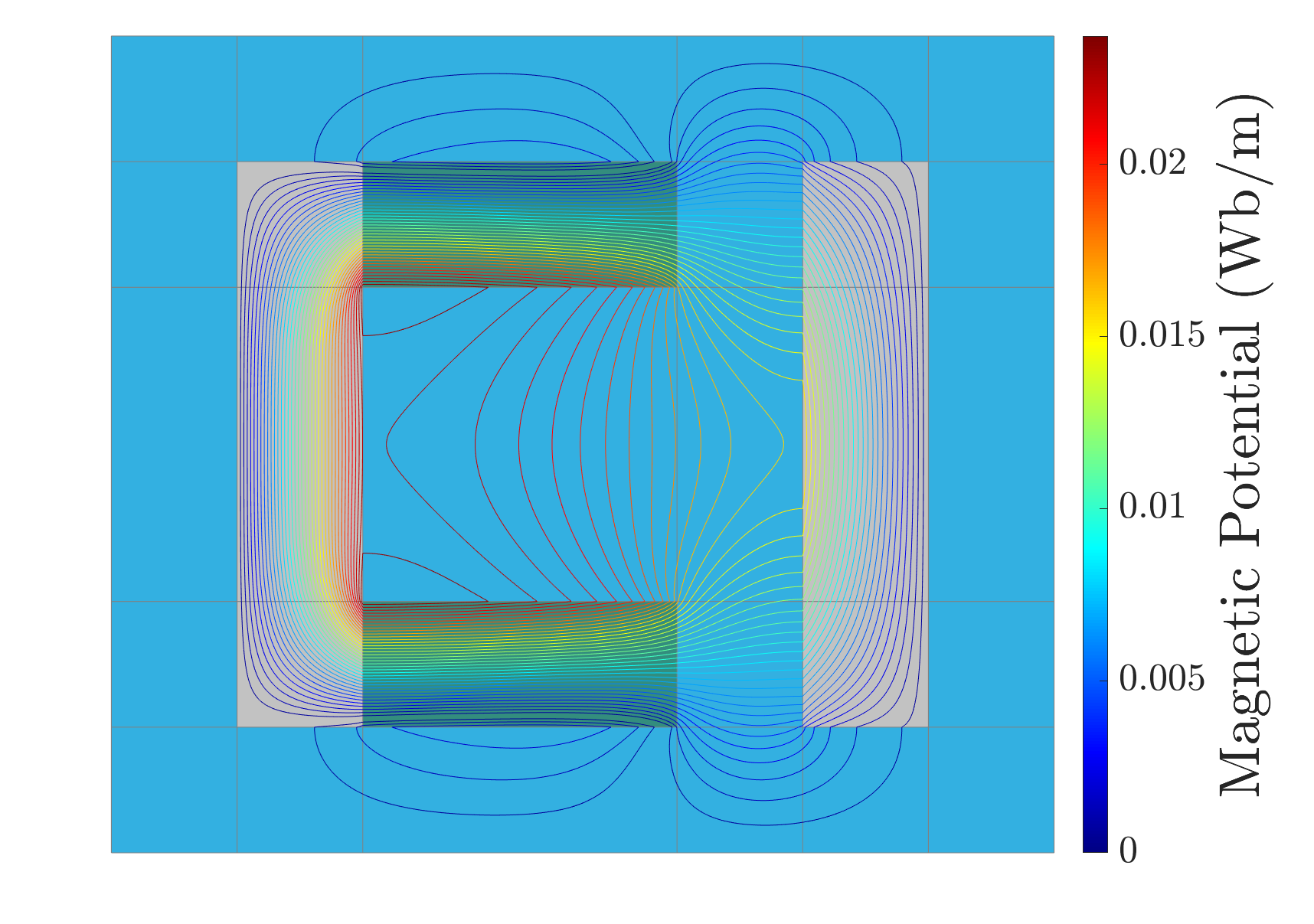}
    \end{minipage}
    \hfil
    \begin{minipage}{.49\textwidth}
        \includegraphics[scale=0.24]{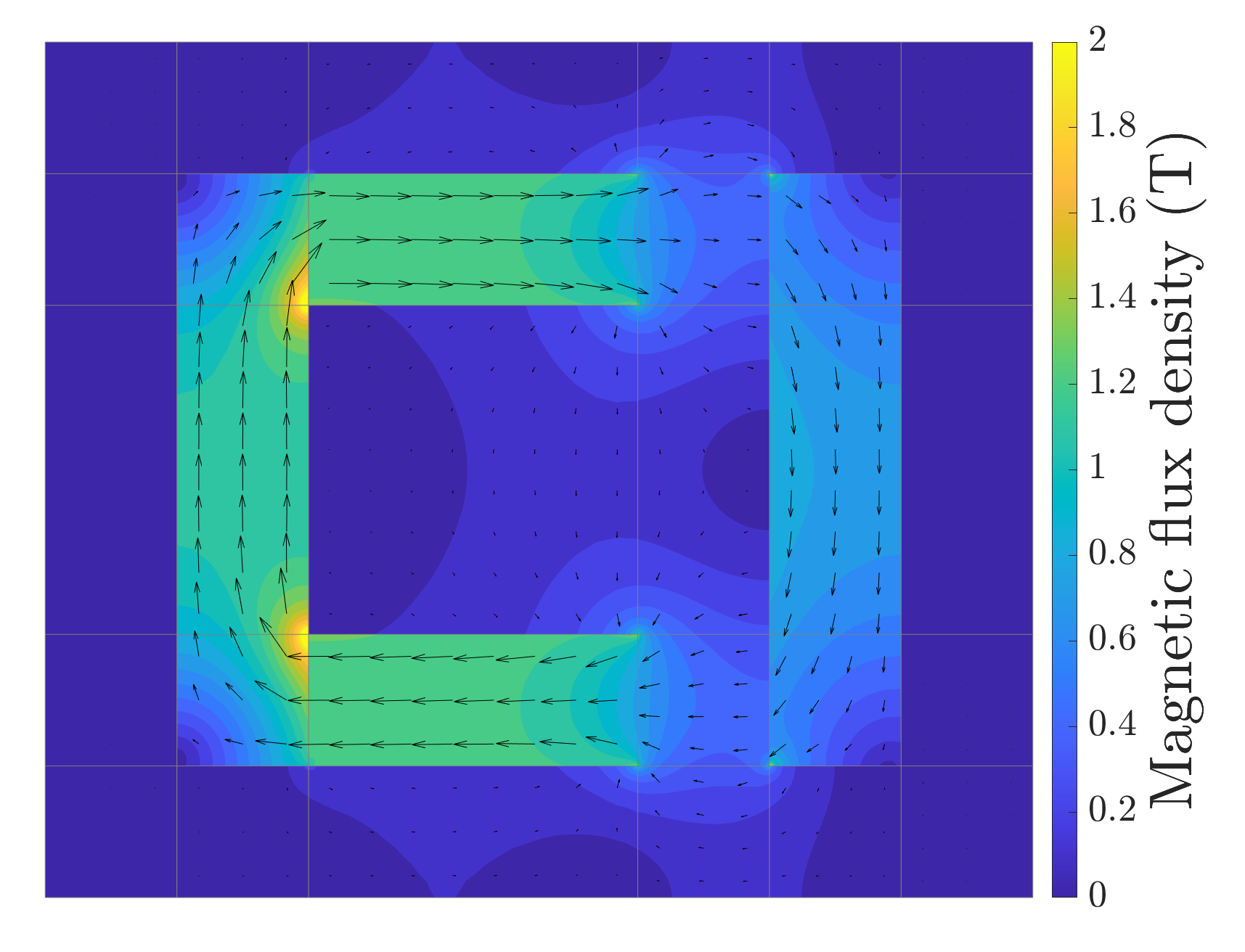}
    \end{minipage}
    \caption{Left: magnetic potential lines. Right: magnetic field density.}
    \label{fig:post}
\end{figure}

The boundary conditions are homogeneous Dirichlet conditions, i.~e.~, a flux wall, to ensure that no magnetic field exists outside our domain, see Fig.~\ref{fig:post} left. Then, the magnetic flux density can be derived by the definition of the magnetic vector potential, as shown in Fig.~\ref{fig:post} right.

\begin{figure}
    \centering \includegraphics[scale=0.245]{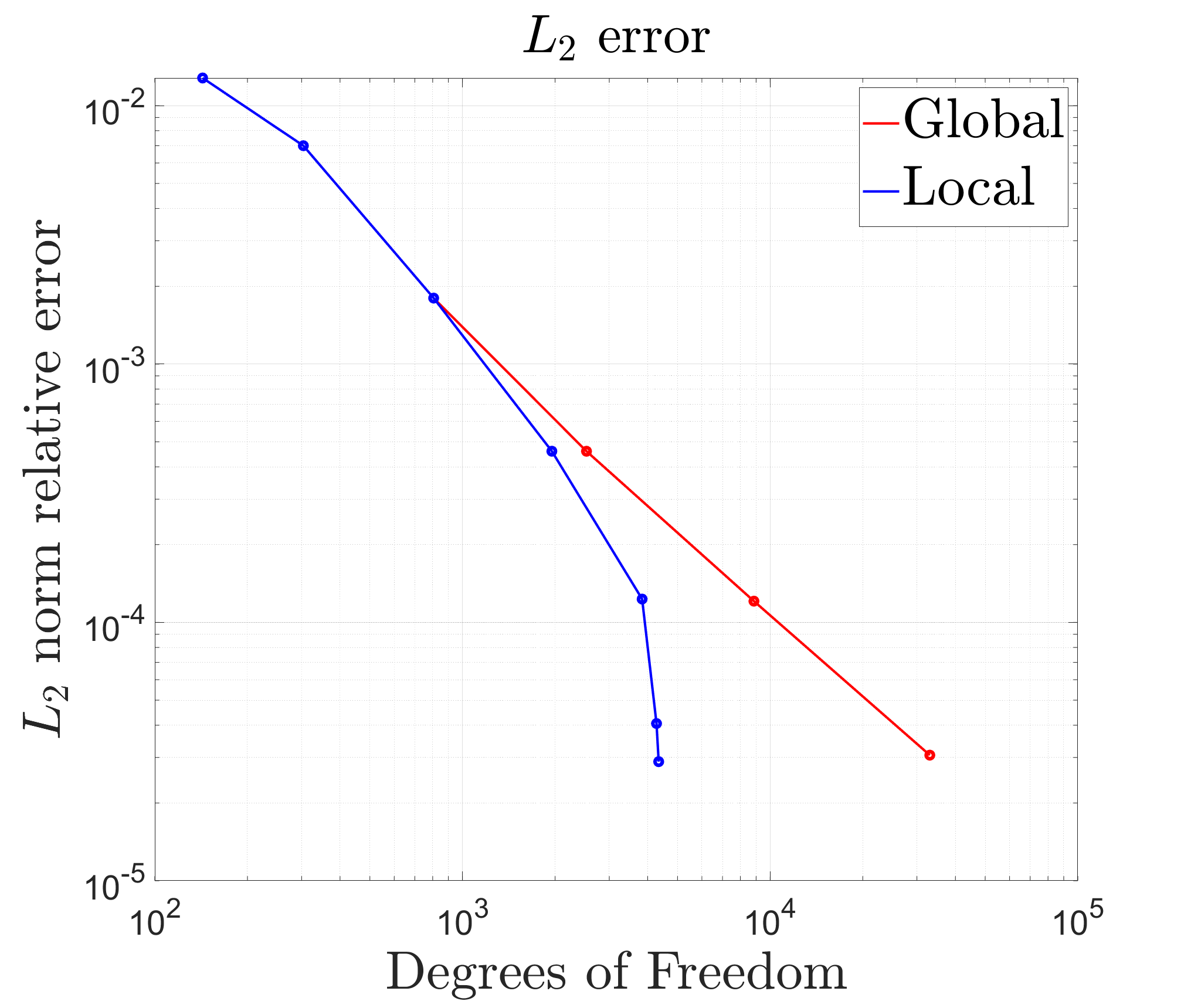}
    \caption{$L_2$ norm relative error, local and global refinement.}
    \label{fig:error}
\end{figure}

 Finally, the efficiency of the multi-level B\'{e}zier extraction for local refinement can be described by the relative $L_2$-error approximation between uniform and local refinement, as shown in Fig.~\ref{fig:error}. As expected, the local refinement converges faster with fewer degrees of freedom, but at the same time, this verifies the effectiveness of multi-level B\'{e}zier extraction.

\section{Conclusions}
We discussed the extension of the B\'{e}zier extraction technique to truncated hierarchical B-splines (THB-splines). We demonstrated the behavior of this approach in the context of 2D magnetostatic simulation. The multi-level B\'{e}zier extraction can be compatible with other hierarchical spline spaces and is not restricted to THB-splines. The example of multipatch domains demonstrates that multi-level B\'{e}zier extraction applies to complex geometries, where the calculation of B\'{e}zier elements suffices to simulate over hierarchical spaces. Since the assembly is performed element-by-element with a subsequent multiplication of the extraction operators, integrating THB-splines into other B\'{e}zier element-based isogeometric analysis (IGA) codes is straightforward. Additionally, the use of standard FE codes is simplified significantly. The approximation error drives the refinement process. The error was computed using an error estimator based on consecutive levels of refinement in one example, while the other utilized a reference solution.

\section{Ackowledgments}
The work is supported by the joint DFG/FWF Collaborative Research Centre CREATOR (DFG: Project-ID 492661287/TRR 361; FWF: 10.55776/F90) at TU Darmstadt, TU Graz and JKU Linz.

\end{document}